**Altermagnetism: Exploring New Frontiers in Magnetism and Spintronics**

*Ling Bai, Wanxiang Feng*, Siyuan Liu, Libor Šmejkal, Yuriy Mokrousov, Yugui Yao**

Ling Bai[1,2], Wanxiang Feng[1,2], Siyuan Liu[1,2]
[1]Centre for Quantum Physics, Key Laboratory of Advanced Optoelectronic Quantum Architecture and Measurement (MOE), School of Physics, Beijing Institute of Technology, Beijing 100081, China
[2]Beijing Key Lab of Nanophotonics and Ultrafine Optoelectronic Systems, School of Physics, Beijing Institute of Technology, Beijing 100081, China
E-mail: wxfeng@bit.edu.cn

Libor Šmejkal[3,4,5]
[3]Max Planck Institute for the Physics of Complex Systems, Nöthnitzer Str. 38, 01187 Dresden, Germany
[4]Institute of Physics, Johannes Gutenberg University Mainz, 55099 Mainz, Germany
[5]Institute of Physics, Czech Academy of Sciences, Cukrovarnická 10, 162 00 Praha 6, Czech Republic

Yuriy Mokrousov[4,6]
[4]Institute of Physics, Johannes Gutenberg University Mainz, 55099 Mainz, Germany
[6]Peter Grünberg Institut and Institute for Advanced Simulation, Forschungszentrum Jülich and JARA, 52425 Jülich, Germany

Yugui Yao[1,2]
[1]Centre for Quantum Physics, Key Laboratory of Advanced Optoelectronic Quantum Architecture and Measurement (MOE), School of Physics, Beijing Institute of Technology, Beijing 100081, China
[2]Beijing Key Lab of Nanophotonics and Ultrafine Optoelectronic Systems, School of Physics, Beijing Institute of Technology, Beijing 100081, China
E-mail: ygyao@bit.edu.cn

Recent developments have introduced a groundbreaking form of collinear magnetism known as “altermagnetism”. This emerging magnetic phase is characterized by robust time-reversal symmetry breaking, antiparallel magnetic order, and alternating spin-splitting band structures, yet it exhibits vanishing net magnetization constrained by symmetry. Altermagnetism uniquely integrates traits previously considered mutually exclusive to conventional collinear ferromagnetism and antiferromagnetism, thereby facilitating phenomena and functionalities previously not achievable within these traditional categories of magnetism. Initially proposed theoretically, the existence of the altermagnetic phase has since been corroborated by a range of experimental studies, which have confirmed its unique properties and potential for applications. This review explores the rapidly expanding research on altermagnets, emphasizing the novel physical phenomena they manifest, methodologies for inducing altermagnetism, and promising altermagnetic materials. The goal of this review is to furnish readers with a comprehensive overview of altermagnetism and to inspire further innovative studies on altermagnetic materials which could potentially revolutionize applications in technology and materials science.

## 1. Introduction

Magnetism, a fundamental and expansive area within condensed-matter physics, plays a crucial role in advancing technology. Historically, this field has primarily focused on two magnetic phases: ferromagnetism and antiferromagnetism. Ferromagnets (FMs), known for their spin polarization that mirrors the macroscopic magnetization, facilitate a myriad of time-reversal ($\mathcal{T}$) symmetry-breaking responses and have therefore been the subject of extensive research and widespread applications.[1-4] Nevertheless, the inherent properties of FMs present significant challenges for the performance of ferromagnetic spintronic devices, particularly affecting their stability, scalability, and dynamic responses. These limitations render them suboptimal for incorporation into advanced electronic systems. Specifically, the field-sensitivity of ferromagnetic order exposes these materials to external magnetic disturbances, substantial stray magnetic fields compromise the potential for high-density integration, and the conventional ferromagnetic precession frequencies, typically on the order of GHz, constrain the operational speed of these devices. In conventional antiferromagnets (AFMs), the perfectly compensated antiparallel magnetization results in zero net magnetic moment, which confers significant robustness against magnetic disturbances and eliminates stray magnetic fields. Furthermore, due to antiferromagnetic exchange interactions, the precession frequencies of AFMs can extend up to the THz range, thus opening up a new frontier in antiferromagnetic spintronics.[5-8] However, the compensated magnetization of AFMs is often perceived as leading to uncontrollable antiferromagnetic order, weak magnetic signal, and very intricate magnetoelectronic responses. To overcome the limitations of collinear ferromagnetic and antiferromagnetic materials in spintronics, scientists have been exploring new materials that can combine the advantages of both ferromagnetism and antiferromagnetism. A new magnetic phase—altermagnetism—has consequently emerged.

With compensated antiparallel magnetic order, altermagnetic band structure breaks $\mathcal{T}$ symmetry and possesses characteristic nonrelativistic spin-splitting. Indication of unconventional magnetism can be tracked back to reports of anomalous effects by various research groups.[9-21] The anomalous effects have been referred to by several names, such as crystal Hall effect,[9] AFM-induced spin splitting,[12] staggered spin-momentum interaction,[14] spin-splitter currents,[19] unconventional antiferromagnetism and valley-dependent spin-momentum interactions,[21] C-paired spin-momentum locking,[17] unconventional magnetism and anti-Kramers nodal surfaces.[15] Shortly after, it was recognized that these anomalous properties[9-21] and other results[22-23] are characteristics of an unconventional magnetic class. This unconventional magnetic class, identified by symmetries by Šmejkal and colleagues, exhibits alternating spin polarization with *d*-, *g*-, or *i*-wave symmetry (as shown in **Figure 1**) in both direct and reciprocal space; therefore, it has been aptly named as "altermagnetism", a term that has gained wide acceptance.[24] Although it has, by symmetry, no net magnetization, unlike conventional antiferromagnetism, the sublattices with opposite spins are connected by a real-space rotational transformation—either proper or improper, and either symmorphic or non-symmorphic—rather than by translation or inversion.[24] As a result, while altermagnets (AMs) share certain key properties with AFMs, they demonstrate even more similarities with FMs due to the alternating spin-splitting of the bands in the absence of spin-orbit coupling (SOC).[25] Numerous studies have investigated the nature of AMs both theoretically and experimentally,[14, 21, 26-38] developing a Landau theory for AMs,[39] constructing models,[21, 24, 40-46] and connecting altermagnetism with superconductivity,[26, 47-56] topological phenomena[40, 57] and magnetic multipoles.[40, 58] Moreover, certain effects such as anomalous Hall effect can be found also in noncollinear systems as discussed, for instance, in reviews[7, 59] and work by Cheong *et al*.[60] Additionally, unconventional odd parity-wave magnets were proposed[61] which are noncollinear and, unlike altermagnets, exhibit time-reversal symmetric exchange spin split bands.[61-62] However, this review focuses solely on collinear systems and AMs.

We note that there have been several comprehensive reviews covering topics such as antiferromagnetic spintronics, anomalous Hall antiferromagnets, and the emerging field of altermagnetism.[8, 26, 59, 63] Some of these reviews were published at a time when research into AMs was still in its early stages and predominantly theoretical,[26] while others concentrated on specific physical properties or considered complex magnetic configurations, such as noncollinear or noncoplanar antiferromagnetism, in which the coverage of AMs was notably limited.[8, 59, 63] Here, we focus on the recent theoretical and experimental progress in AMs, highlighting canonical examples such as MnTe,[24, 31, 64-66] CrSb[24, 35, 67-71] and $RuO_2$[24, 32, 72] to offer a detailed exploration of novel physical phenomena associated with altermagnetism.

In this review, we first introduce several exotic physical phenomena manifested in AMs, including lifted Kramers degeneracy, anomalous and spin transport properties, magneto-optical effects and chiral magnons. Then we discuss various strategies for inducing altermagnetism from conventional FMs and AFMs, which broaden the application scope of traditional magnetic materials. In addition to artificial AMs, naturally occurring AMs also play a significant role; therefore, we compile a list of potential two-dimensional (2D) and three-dimensional (3D) altermagnetic candidates found in nature. Although our understanding of altermagnetism remains in its nascent stages, the field has already become a rich area for theoretical exploration and a promising avenue for the development of spintronic devices.

## 2. Exotic physical phenomena

AMs have demonstrated a series of phenomena that were previously considered exclusive to FMs, including nonrelativistic lifted Kramers degeneracy, anomalous Hall/Nernst effect, nonrelativistic spin (polarized) currents, and the magneto-optical Kerr effect. Moreover, AMs also display unique physical phenomena, e.g. the emergence of chiral magnons. In this section, we will primarily focus on lifted Kramers degeneracy, anomalous and spin transport properties, magneto-optical effects and chiral magnons in AMs.

### 2.1. Lifted Kramers degeneracy

We start by briefly recalling three kinds of space groups. Symmetry, one of the central concepts in modern physics, plays a significant role in physical properties, enabling physicists to make *a priori* judgments without knowledge of certain microscopic details. More than one hundred years ago, based on group theory, the symmetries of 3D crystals were classified into 230 space groups. However, these 230 space groups are only applicable to nonmagnetic cases. For magnetic systems, an additional degree of freedom, associated with spin, arises. The spin magnetic moments are also periodically arranged in spin space, forming a 3D pseudovector field. This pseudovector field exhibits even parity under space inversion ($\mathcal{P}$) symmetry and odd parity under $\mathcal{T}$ symmetry. Initially, the combination of $\mathcal{T}$ symmetry with the aforementioned 230 space groups resulted in 1651 magnetic space groups. In these magnetic space groups, the rotation of spin is coupled with spatial operations, meaning the orientation of spin may change under spatial operations. Therefore, when considering SOC, it is essential to take magnetic space groups into account in symmetry analysis of magnetic materials. Later in the 1960's, Brinkman and Elliott proposed that in many cases, it was sufficient to include only Heisenberg exchange and anisotropy field in a spin Hamiltonian, while neglecting SOC.[73] Such groups, which usually have more symmetries, were introduced as "spin-space groups". In a spin-space group, the spins are rotated independently of the lattice. Therefore, the spin-symmetry operation can be represented as $[\mathcal{R}_s||\mathcal{R}_l]$, where the transformation on the left (right) of the double vertical bar acts solely in spin space (spatial space). Although in magnetism, the key features of the electronic structure and basic characteristics can be explained without referring to nonrelativistic physics, people are used to describing the symmetry properties of magnetic systems in terms of the magnetic space groups due to the complexity of spin-space group. Only in recent years has research involving spin-space groups matured significantly,[15, 24, 26, 73-79] with the most noteworthy development being the AMs with lifted Kramers degeneracy.

Kramers degeneracy theorem[80-82] states that any system possessing combined $\mathcal{PT}$ symmetry ensures at least double spin degeneracy at arbitrary wave vectors. Typically, the breaking of Kramers degeneracy is considered to originate from two fundamental mechanisms: relativistic and nonrelativistic. Here we focus on the latter. For all collinear magnets, the spin-only symmetry $[\mathcal{C}_\infty||E]$ guarantees that spin is a good quantum number and is momentum ($k$)-independent, where $\mathcal{C}_\infty$ denotes arbitrary rotations of spin around the common axis of spins, and $E$ is the identity transformation in real space. Therefore, the energy band can be represented by $E(k,s)$, where $s$ denotes spin. Another common spin-only symmetry in collinear magnets is $\bar{\mathcal{C}}_2$, representing a twofold rotation perpendicular to the collinear spin axis, followed by spin-space inversion, which we denote also as $[\mathcal{C}_2||\mathcal{T}]$. This symmetry ($\bar{\mathcal{C}}_2$) together with collinear symmetry $\mathcal{C}_\infty$ implies nonrelativistic inversion symmetry ($\mathcal{P}$) of the band structure,[61] as it transforms $E(k,s)$ to $E(-k,s)$, and thus enforces $E(k,s)=E(-k,s)$. Consequently, the bands exhibit even parity in momentum around the Γ point. From the analysis above, we observe that for any type of collinear magnets, the nonrelativistic bands remain invariant under $\mathcal{P}$ symmetry, regardless of whether the system possesses real-space inversion symmetry. This observation means that if a collinear magnet has $[\mathcal{C}_2||\mathcal{P}]$ symmetry, the spin band is Kramers degenerate across the entire Brillouin zone. This degeneracy emerges because the combination $[\mathcal{C}_2||\mathcal{P}][\mathcal{C}_2||\mathcal{T}]$ enforces $E(k,s)=E(k,-s)$, which is equivalent to combined $\mathcal{PT}$. In addition, the symmetry denoted by $[\mathcal{C}_2||E]/[\mathcal{C}_2||\tau]$ ($\tau$ means translational symmetries) also enforces $E(k,s)=E(k,-s)$. Given that all nonmagnets have $[\mathcal{C}_2||E]$ symmetry, Kramers degeneracy is present in all nonmagnetic systems in the nonrelativistic limit. In FMs, if the spin moments of all magnetic atoms are flipped (under spin-only symmetry), there is no symmetry that acts solely on the lattice to recover the initial magnetic configuration. Consequently, Kramers degeneracy is broken in FMs, resulting in uniform Zeeman splitting between the opposite-spin bands. For systems formally spin-compensated by symmetry, there are two types of magnetic order: conventional AFMs, which include the symmetries of $[\mathcal{C}_2||\mathcal{P}][\mathcal{C}_2||\mathcal{T}]$ and/or $[\mathcal{C}_2||\tau]$, and AMs, which lack these two symmetries. It is obviously that the former is spin degenerate, and the latter is spin nondegenerate. The nonrelativistic band structures of these three types of collinear magnets are depicted in **Figure 1A**.[24] The spin rotational $\mathcal{C}_2$ symmetry, along with rotational symmetry connecting opposite-spin sublattice, causes the electronic band structures of AMs to show *d*-, *g*- or *i*-wave anisotropy (alternating spin polarizations) and characteristics of spin-momentum locking. These phenomena have been comprehensively analyzed in ref. [24], and we selectively show the planar even-parity wave form of altermagnetism in Figure 1B.

Noncollinear systems can also show spin splitting with complex patterns,[83] but it is out of the scope of this review.

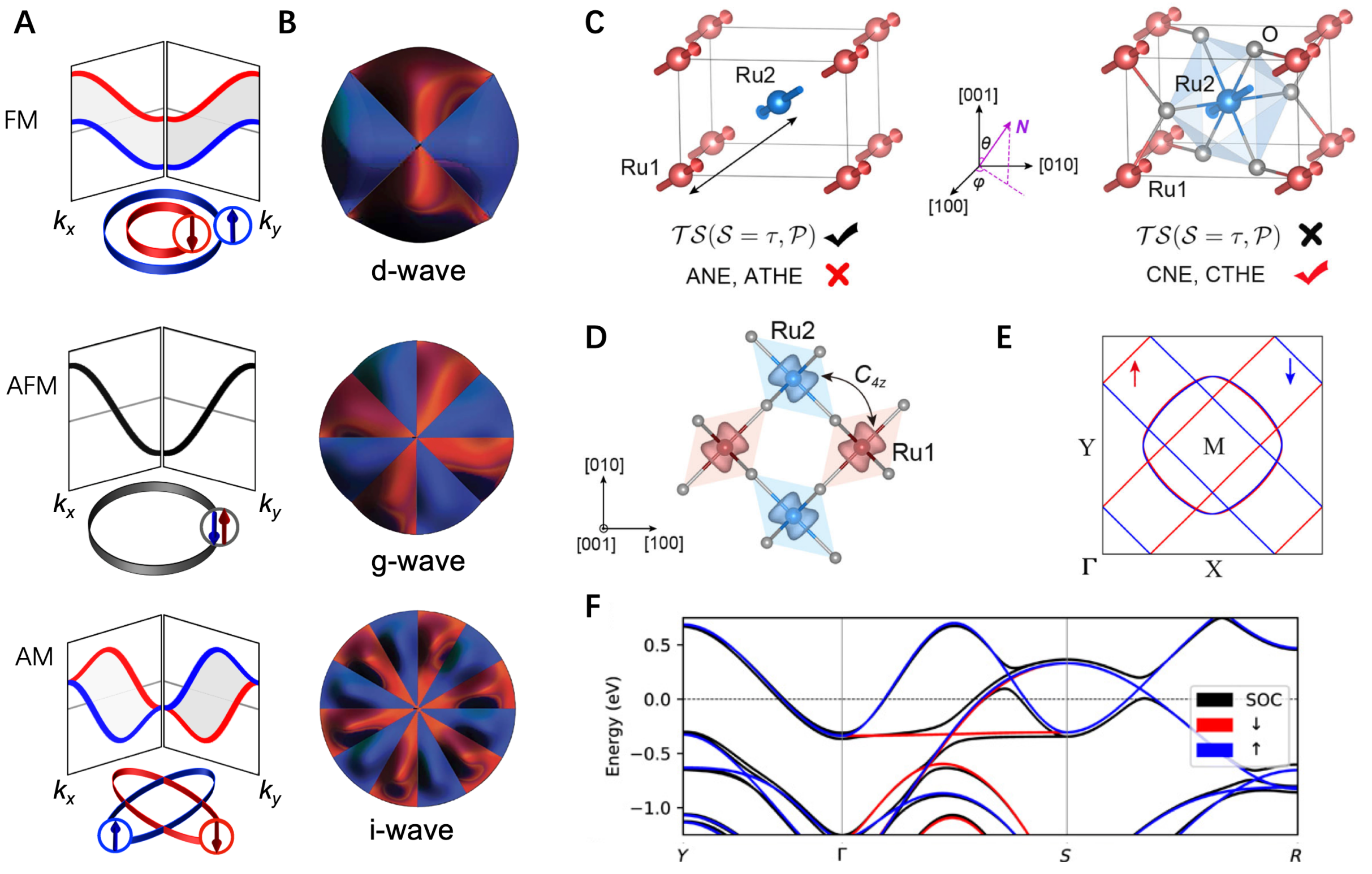

**Figure 1**. A) Schematic diagrams of band structures and corresponding energy iso-surfaces of the three nonrelativistic collinear magnetic phases. B) The planar even-parity wave form of altermagnetism. Reproduced under the terms of the CC-BY 4.0 license.[24] Copyright 2022, The Authors, published by American Physical Society. C) Magnetic unit cell of $RuO_2$ without and with O atoms. Two Ru atoms with antiparallel spin magnetic moments are represented by red and blue balls, and the nonmagnetic O atoms are depicted by gray balls. D) Magnetization density related by fourfold crystal rotational symmetry. E) Reciprocal space Fermi surface cut at wave vector $k_z = 0$ calculated without spin-orbit coupling. Reproduced with permission.[30] Copyright 2024, American Physical Society. F) Band structure of $RuO_2$ with (black) and without (red and blue) spin-orbit coupling. Reproduced under the terms of the CC-BY-NC 4.0 license.[9] Copyright 2020, The Authors, Published by American Association for the Advancement of Science.

Next, we use the most investigated altermagnetic candidate $RuO_2$ as a case study to specifically illustrate the d-wave anisotropic spin-momentum locking observed in its band structure. Although the magnetic ground state of single crystalline $RuO_2$ remains controversial,[84-87] assuming hypothetical altermagnetic order for $RuO_2$ is justifiable, given our primary interest in exploring the interplay between altermagnetism and band structure. As shown in Figure 1C,[30] crystals composed solely of the magnetic atom Ru possess $[\mathcal{C}_2||\tau]$ symmetry, and thus belong to the category of conventional collinear AFMs (left panel). The arrangement of the nonmagnetic oxygen atoms leads to the breaking of this symmetry, resulting in altermagnetic order (right panel). Therefore, the positioning of nonmagnetic atoms significantly influences the magnetic phase of a material. The spin-space group of $RuO_2$ incorporates the following symmetry operations

$$[E||H] + [\mathcal{C}_2||G - H] = [E||H] + [\mathcal{C}_2||AH] \quad (1)$$

where $G$ represents the space group of $RuO_2$ and $H$ is a halving subgroup of G. $H$ contains $E, \mathcal{P}, \mathcal{C}_{2z}, \mathcal{C}_{2[110]}, \mathcal{C}_{2[\bar{1}10]}, \mathcal{M}_z, \mathcal{M}_{[110]}, \mathcal{M}_{[\bar{1}10]}$ and $G - H$ includes $\tau\{\mathcal{C}_{2x}, \mathcal{C}_{2y}, \pm\mathcal{C}_{4z}^{\pm}, \mathcal{M}_x, \mathcal{M}_y\}$, with $\tau$ representing half a unit cell translation and $A$ is operation from $G - H$ such as $\tau\mathcal{C}_{4z}^{+}$. We note that $[E||H]$ are transformations that interchange atoms within only one of the two spin sublattices. These symmetries are crucial in determining the characteristic anisotropy of the spin density on the opposite-spin sublattices (Figure 1D).[30] $[\mathcal{C}_2||AH]$ comprises symmetries that interchange atoms between opposite-spin sublattices and guarantees zero net magnetization. It is noted that $AH$ does not contain $\mathcal{P}$, indicating the broken $\mathcal{T}$ symmetry in the band structure, i.e., $E(k,s) \neq E(-k,-s)$. Moreover, for momentum whose little group does not contain $AH$ elements, such as $\Gamma - S$ path (Figure 1F), the spin-up and spin-down bands are nondegenerate, i.e., $E(k,s) \neq E(k,-s)$.[9] The magnitude of the spin splitting can reach up to eV scale, comparable to that observed in FMs and even larger than that induced by relativistic SOC.[26, 65, 72] $[\mathcal{C}_2||\mathcal{M}_x]$ and $[\mathcal{C}_2||\mathcal{M}_y]$ (disregarding translational symmetry for simplicity) respectively transform $E(k,s)$ as $[\mathcal{C}_2||\mathcal{M}_x]E(k_x,k_y,k_z,s) = E(-k_x,k_y,k_z,-s)$ and $[\mathcal{C}_2||\mathcal{M}_y]E(k_x,k_y,k_z,s) = E\big(k_x,-k_y,k_z,-s\big)$. These two transformations enforce the spin degeneracy on the $k_x = 0,\pi$ and $k_y = 0,\pi$ planes. In addition, $[\mathcal{C}_2||\mathcal{C}_{4z}^{+}]$ transforms $E(k,s)$ as $[\mathcal{C}_2||\mathcal{C}_{4z}^{+}]\, E(k_x,k_y,k_z,s) = E\big(k_y,-k_x,k_z,-s\big)$. This transformation makes the spin splitting of bands alternate across two perpendicular wave vectors in the $k_z$ plane. These symmetries collectively lead to a planar d-wave spin-momentum locking band structure of $RuO_2$, as shown in Figure 1E.[30]

In **Figure 2A-C**, we present the *ab initio* nonrelativistic band structures of other AMs, such as metallic $FeSb_2$[15], organic κ-(BEDT-TTF)$_2$Cu[N(CN)]Cl (abbreviated as κ-Cl), and insulating $RuF_4$[88], which also illustrate that altermagnetic spin splitting is strongly momentum dependent and the sign of splitting is alternating across the Brillouin zone. Figure 2D illustrate the average energy splitting between opposite-spin bands across a broad pool of altermagnetic candidates, with different color sectors corresponding to different magnetic atoms.[89] We observe that no specific magnetic atoms can be identified consistently as producing particularly sizable spin splitting, and in AMs composed solely of light elements, significant band splitting also occurs.

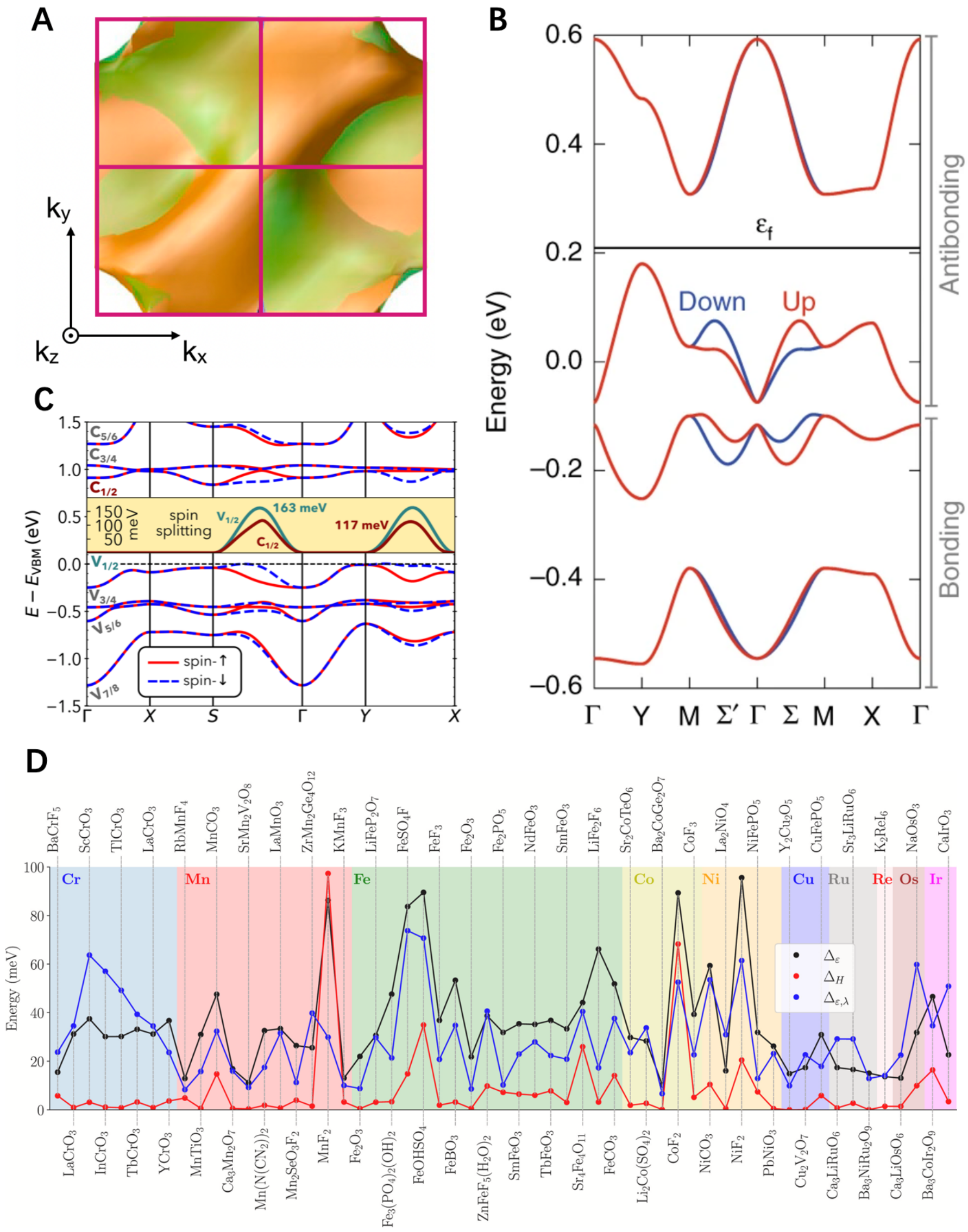

**Figure 2**. A) Top view of the spin-polarized Fermi surface of $FeSb_2$. Reproduced under the terms of the PNAS License.[15]Copyright 2019, The Authors, Published by PNAS. B) Nonrelativistic κ-Cl band structure. Reproduced under the terms of the CC-BY 4.0 license.[90] Copyright 2019, The Authors, Published by Springer Nature. C) Nonrelativistic $RuF_4$ band structure. The inset indicates the spin splitting energies of the two highest occupied energy bands ($V_{1/2}$) and lowest unoccupied energy bands ($C_{1/2}$). Reproduced under the terms of the CC-BY 4.0 license.[88] Copyright 2024, The Authors, Published by IOP Publishing Ltd. D) The average energy splitting between opposite-spin bands across a broad pool of altermagnetic candidates. Reproduced with permission.[89] Copyright 2023, Elsevier.

In addition to band structures derived from first-principles calculations, several experimental observations have also revealed band splitting phenomena in AMs.[31-32, 35, 64-65, 72] Although a recent experiment showed that the magnetic ground state of bulk $RuO_2$ is non-magnetic,[87] a time-reversal symmetry broken band structure of epitaxial $RuO_2$ film was observed by detecting magnetic circular dichroism in angle-resolved photoemission spectra (**Figure 3A**).[72] Conducted at the [001] orientation of the Ru moments, the measurements reveal no detectable remnant net magnetization, indicating unconventional origin of the signal. Most detailed experimental evidence of theoretically predicted[24] g-wave altermagnetic spin splitting was reported in studies on MnTe, [31, 64-66] see Figure 3B. The studies reported momentum-dependent lifting of Kramers degeneracy[31, 64-66], quadratic spin splitting form around Γ point, [31] and temperature dependence of the splitting.[64, 66] Furthermore, Reimers *et al.* employed spin-integrated soft X-ray angular-resolved photoelectron spectroscopy (SX-ARPES) to investigate the band structure of epitaxial CrSb thin films.[35] As shown in Figure 3C, their observations of band dispersions, consistent with theoretical calculations that predict spin splitting, strongly support the existence of altermagnetic band structure in CrSb. The measured spin splitting, with a significant magnitude of approximately 0.6 eV below the Fermi energy, highlights the potential applications in spintronics.

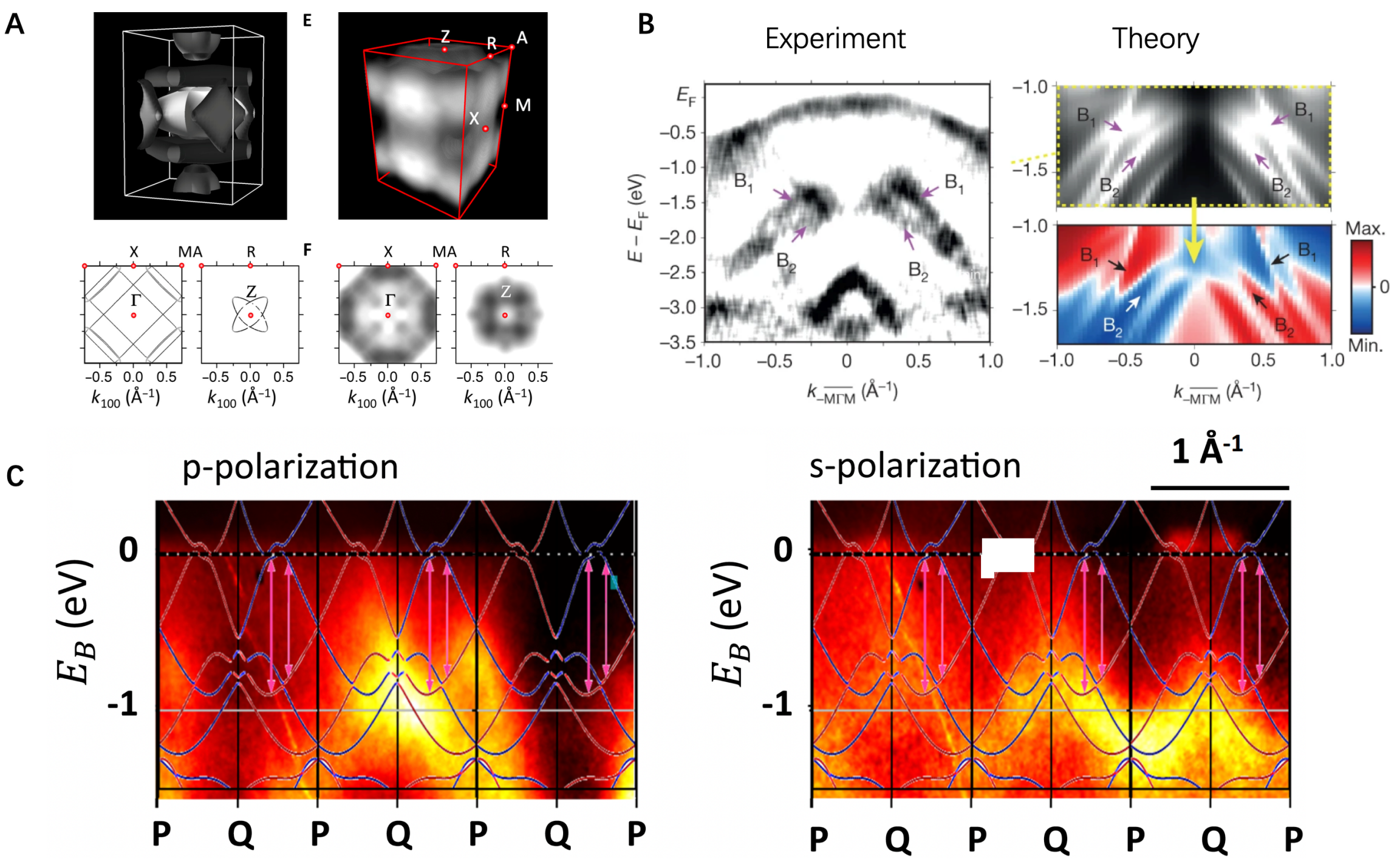

**Figure 3**. A) Left panel: Calculated 3D Fermi surface and Fermi surface cut for $RuO_2$. Right panel: Experimental Fermi energy intensity distribution and the same Fermi surface cut for $RuO_2$. Reproduced under the terms of the CC-BY 4.0 license.[72] Copyright 2024, The Authors, Published by American Association for the Advancement of Science. B) Measured angular-resolved photoelectron spectroscopy (ARPES) band dispersion at $k_z = 0.35$ Å$^{-1}$ along the $\bar{\Gamma} - \bar{M}$ path and corresponding spin-resolved APRES simulation for α-MnTe. Blue and red colors indicate spin polarization. Reproduced under the terms of the CC-BY 4.0 license.[31] Copyright 2024, The Authors, Published by Springer Nature. C) The spin integrated soft X-ray angular-resolved photoelectron spectroscopy (SX-ARPES) intensity, along with the superimposed spin-resolved band structure, is shown for CrSb along the high-symmetry P-Q path using both p-polarized and s-polarized photons. Reproduced under the terms of the CC-BY 4.0 license.[35] Copyright 2024, The Authors, Published by Springer Nature.

The spin splitting of altermagnetic bands makes it possible to realize numerous spin-dependent phenomena, such as anomalous Hall effect, magneto-optical Kerr effect, unconventional longitudinal and transverse spin currents, giant magnetoresistance (GMR), and tunneling magnetoresistance (TMR) effects and their thermal and magnonic counterparts. We will now discuss these effects highlighting their potential for faster and scalable spintronics devices.

## 2.2. Anomalous and spin transport properties

### 2.2.1. Anomalous Hall, Anomalous Nernst, and Anomalous thermal Hall effects

The anomalous Hall effect (AHE)[91] is characterized by a transverse voltage drop that occurs in response to a longitudinal charge current, even in the absence of an applied magnetic field. The presence of AHE requires broken $\mathcal{PT}$ and $\mathcal{T}\tau$ symmetry and therefore, the AHE research was previously limited to FMs and complex noncollinear AFMs,[92] as the AHE was not expected to occur in conventional AFMs. AMs naturally meet this requirement, thus becoming candidate materials for generating low-dissipation Hall currents, and indeed the emergence of AHE in AMs has been demonstrated both theoretically and experimentally.[9, 14, 30, 93-102] In 2020, the AHE was theoretically predicted in $RuO_2$ and $CoNb_3S_6$, and the corresponding term “crystal Hall effect” (CHE) was introduced to mark its unconventional properties when compared with conventional anomalous Hall effect arising from magnetization.[9] The CHE arises from unusual symmetry breaking and corresponding order parameters which originate from combination of magnetic and crystallographic order. In the case of altermagnetic $RuO_2$ the magnetic sites have anisotropic crystallographic environment. The symmetry breaking results in a strong angular dependence of the Hall conductivity on the order parameter (akin to crystalline anisotropic magnetoresistance) – e.g. the effect can be completely switched-off by rotating spin quantization axis. The calculated anomalous Hall conductivity of $RuO_2$ with the Néel vector $n$ along [100] as a function of canting angle is shown in **Figure 4A**.[9] The Néel vector $n$ is defined as $n = (m_A - m_B)/2$, where $m_{A/B}$ are the magnetizations of the opposite-spin A and B sublattices. When considering SOC, a small net moment $m = m_A + m_B$ is generated by the Dzyaloshinskii-Moriya interaction (DMI), thus inducing a canting angle between magnetizations of A and B sublattices, resulting in weak ferromagnetism. It is essential to note that the emergence of this weak ferromagnetism is crucial for the electrical manipulation of the altermagnetic order.[103] Therefore, the anomalous Hall conductivity component $\sigma_{xz}$ can be separated into $\sigma_{xz}^{CHE}$ (even in $m$) and $\sigma_{xz}^{AHE}$ (odd in $m$), which represents the contributions induced by altermagnetic order and the small ferromagnetic moment, respectively. We note that the $\sigma_{xz}$ is entirely induced by the altermagnetic order when the canting angle is 0°. Furthermore, $\sigma_{xz}^{CHE}$ dominates the contribution to $\sigma_{xz}$ when the canting angle is less than 10°, suggesting that the effect of the small net magnetic moment on $\sigma_{xz}$ is negligible. Therefore, in the examples to follow, we drop the name "crystal Hall effect", and use commonly accepted "anomalous Hall effect". The experimental observation of AHE in $RuO_2$ is presented in Figure 4C.[95] Note that the measured anomalous Hall conductivity is 1,000 $\Omega^{-1}$ $cm^{-1}$ for $RuO_2$ at 50 T albeit nonzero AHC is picked up already for small fields. The difference to theoretical prediction[9] is likely due to doping effects and the influence of strong magnetic fields.

For $CoNb_3S_6$, the sign of the anomalous Hall conductivity reverses when the crystal chirality is reversed (Figure 4A, right panel).[9] This phenomenon occurs because the anomalous Hall conductivity is odd under $\mathcal{T}$ symmetry, and the two states of crystal chirality are related by $\mathcal{T}$ symmetry. Thus, reversing the Néel vector can also induce a change in sign. Consequently, in AMs, the sign of the anomalous Hall conductivity is determined by the combined effect of the Néel vector and crystal chirality. As illustrated in Figure 4B, the calculated anomalous Hall conductivity for perovskite $CaCrO_3$ is substantial and can be further enhanced by electron or hole doping, which once again underscores the significant impact of doping effects.[98] Additionally, the AHE in epitaxial thin films of $Mn_5Si_3$ and MnTe has also been investigated where spontaneous remanence of the signal was observed.[14, 97, 104] As shown in Figure 4D, it can be observed that the orientation of the Néel vector changes when the external magnetic field is rotated, and the AHE in this material is anisotropic, dependent on the relative orientation of the Néel vector.[102] Therefore, AHE can serve as an effective method for detecting altermagnetic order. It must be mentioned again that not all the states of AMs can generate AHE. Since the anomalous Hall conductivity behaves as an axial vector and is odd under $\mathcal{T}$ symmetry, a nonvanishing AHE is only observed in systems whose symmetries permit such a vector. This implies that the magnetic space group of the system must allow for the existence of a nonzero net magnetic moment. For instance, in $RuO_2$, when the Néel vector is oriented along [001], the magnetic space group is *P4₂'/mnm'*, and all components of anomalous Hall conductivity are prohibited. On the other hand, with the Néel vector aligned along [100], the magnetic space group is *Pnn'm'* and $\sigma_{xz}$ is permitted.

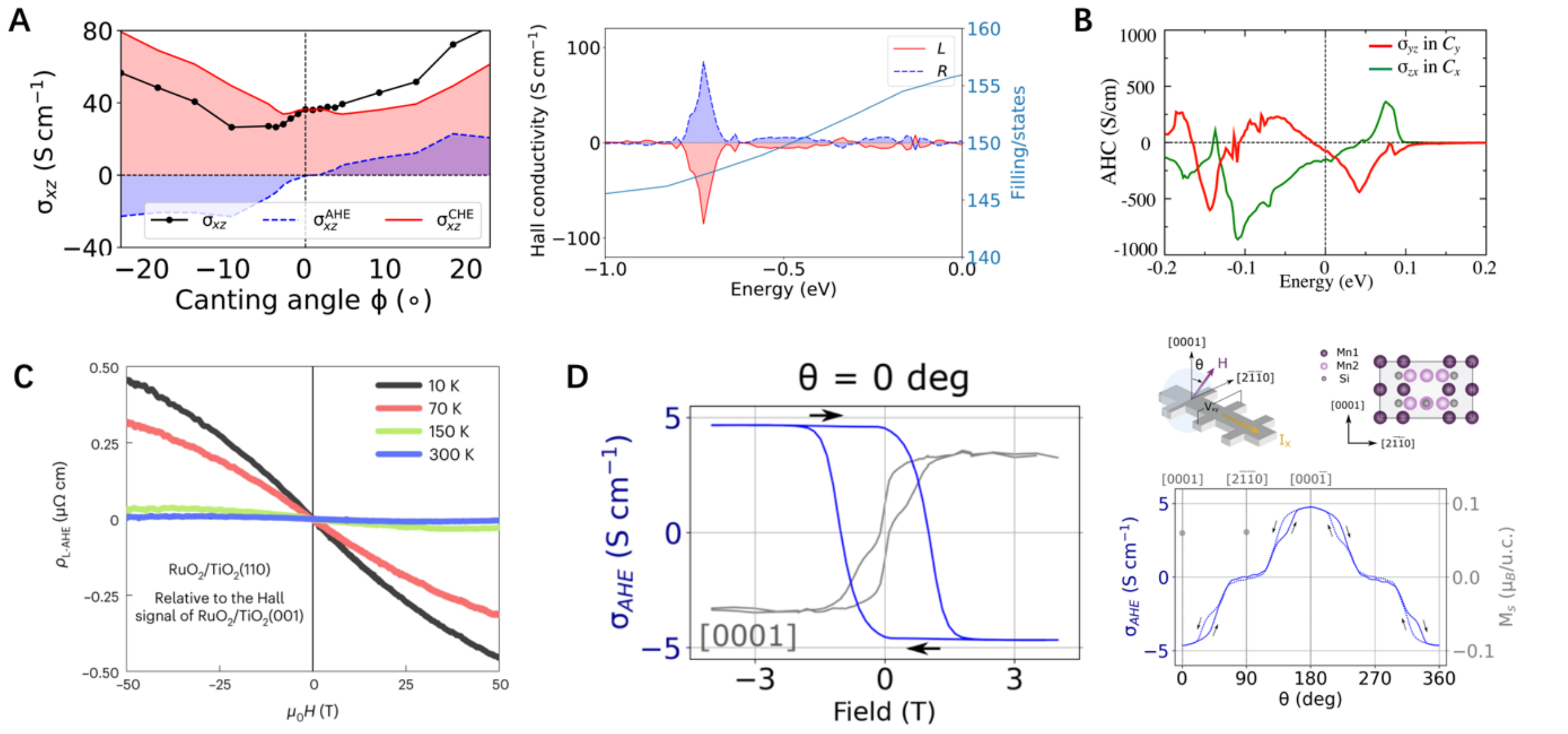

**Figure 4**. A) Left panel: first-principle calculated AHE conductivity in $RuO_2$ as a function of canting angle along with its ferromagnetic and antiferromagnetic components. Right panel: Calculated AHE conductivity in $CoNb_3S_6$ as a function of Fermi level. Reproduced under the terms of the CC-BY-NC 4.0 license.[9] Copyright 2020, The Authors, Published by American Association for the Advancement of Science. B) Calculated AHE conductivity in $CaCrO_3$ as functions of Fermi energy. Reproduced with permission.[98] Copyright 2023, American Physical Society. C) Anomalous Hall resistivity in the (110)-oriented $RuO_2$ film. Reproduced with permission.[95] Copyright 2022, Springer Nature. D) Experimental AHE in $Mn_5Si_3$. Left panel: Experimental AHE conductivity as a function of external field with the orientation $\theta = 0°$. Right panel: Schematic of Hall bar with external field orientation $\theta$ relative to the [0001] direction and measured AHE conductivity as a function of $\theta$. Reproduced with permission.[102] Copyright 2024, American Physical Society.

The anomalous Nernst effect (ANE) and anomalous thermal Hall effect (ATHE) are the thermoelectric and thermal counterparts, respectively, of the AHE. The ANE describes the appearance of a transverse voltage induced by a longitudinal heat current in the absence of an external magnetic field, while the ATHE describes the appearance of a transverse thermal current density under similar conditions. Both phenomena require the same symmetry conditions as AHE. Consequently, all systems that permit the existence of the AHE also allow for the occurrence of the ANE and the ATHE. However, the ANE and the ATHE have been investigated only sporadically in AMs, where the ANE and the ATHE can be termed as crystal Nernst effect and crystal thermal Hall effect, respectively.[30, 33-34] The measured and calculated ANE of $Mn_5Si_3$ is shown in **Figure 5A**.[33] Clear hysteretic behavior and saturation in the transverse thermopower, indicative of a finite ANE, closely correspond to the field dependence of the anomalous Hall conductivity. The measured magnitude of anomalous Nernst conductivity ranges from (0.11 ± 0.08) A/(K·m) at 58 K to (0.015 ± 0.005) A/(K·m) at 216 K, which are in qualitative agreement with the calculated magnitude of 0.25 A/(K·m) for 58 K and 0.12 A/(K·m) for 216 K, respectively. Besides, it has been demonstrated that the anomalous Nernst conductivity undergoes a drastic sixfold enhancement in $Mn_5Si_3$ films with a small amount of Mn doping, especially in $Mn_{5.10}Si_{2.90}$.[34] In addition, Zhou *et al.*[30] analyzed the sources of the large anomalous Nernst conductivity in $RuO_2$ through first-principles calculations, identifying Weyl fermions arising from band crossings, strong spin-flip pseudonodal surfaces, and weak spin-flip ladder transitions as key contributors. Based on this, the total anomalous Hall/Nernst conductivity was decomposed into spin-conserving and spin-flip parts (Figure 5B). The results show that, within the hole doping range of -0.2 to 0.0 eV, the ANE is predominantly driven by the spin-flip processes. Moreover, due to the topological Weyl points, the Wiedemann-Franz law, which reflects the relationship between the anomalous Hall conductivity and anomalous thermal Hall conductivity at low temperatures, is applicable up to 150 K in $RuO_2$, a range significantly broader than what is typically expected of conventional magnets (Figure 5C).[30] Furthermore, Hoyer *et al.* used a toy model of an insulating d-wave AM to investigate the ATHE of magnons, showing that it was strongly dependent on the Néel vector orientation and strain.[105] These works enhance our understanding of anomalous transport behavior associated with alternating spin-splitting band structure in AMs. Given these insights, AMs may play a leading role in the future development of spintronics and spin caloritronics.

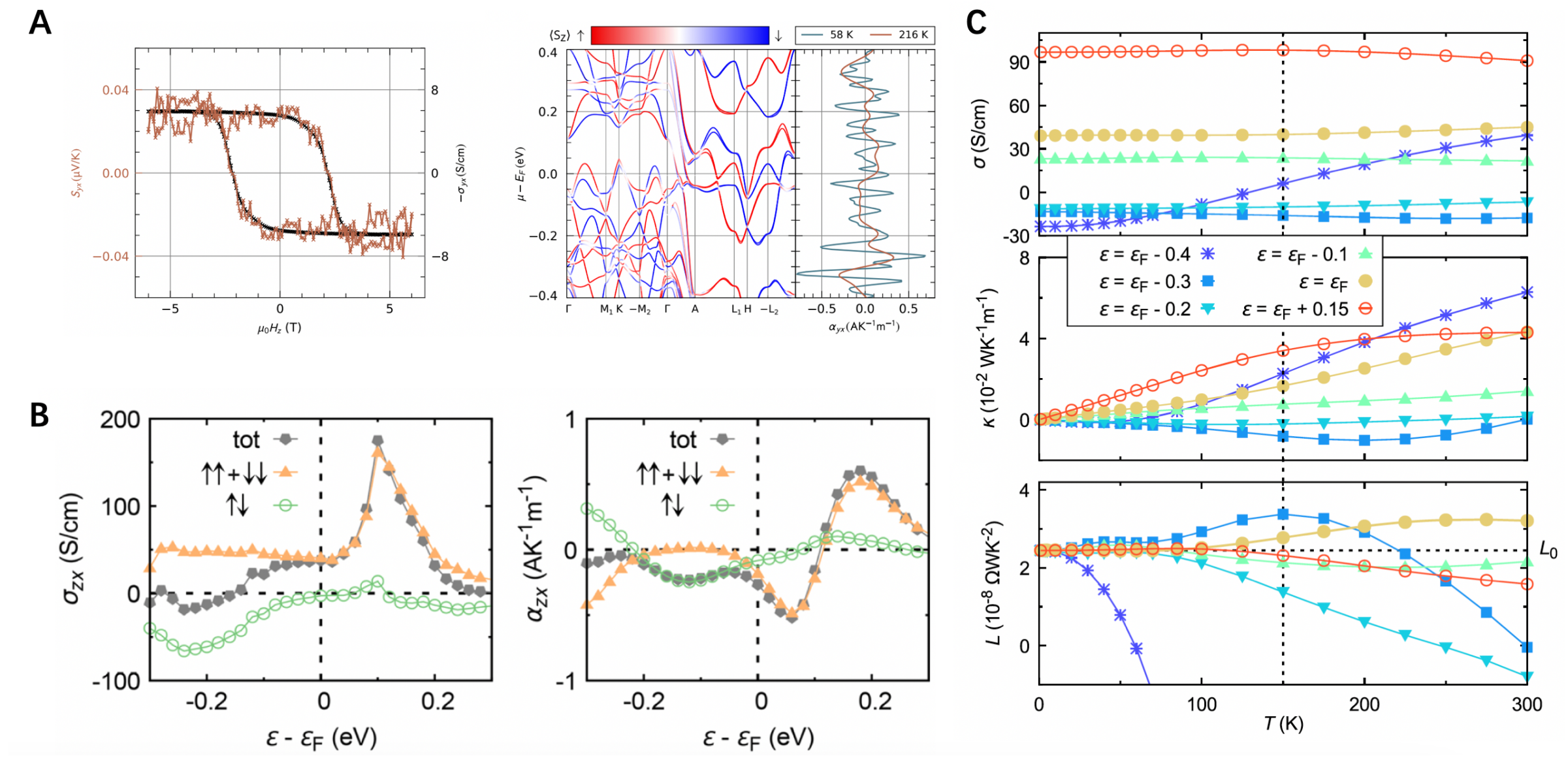

**Figure 5**. A) Transverse Nernst signal (brown) and anomalous Hall conductivity (black) of as a function of applied magnetic field for a sample average temperature of 216 K (left) and *ab initio* calculated band structure and ANE at 58 K and 216 K (right) for $Mn_5Si_3$. Reproduced with permission.[33] Copyright 2024, The Authors, Published by arXiv. B) The calculated total anomalous Hall (left) /Nernst (right) conductivity and their spin-conserved and spin-flip parts of $RuO_2$. C) Anomalous Hall conductivity, anomalous thermal Hall conductivity and anomalous Lorenz ratio as a function of temperature at different Fermi levels. Reproduced with permission.[30] Copyright 2024, American Physical Society.

### 2.2.2. Nonrelativistic spin current

In commercial spin-transfer-torque magnetic random-access memories (STT-MRAMs), the foundational mechanism involves applying a longitudinal spin-polarized current in an out-of-plane direction, moving from a reference to a recording ferromagnetic layer.[106-108] In this context, the longitudinal spin current, originated from ferromagnetic exchange splitting, has high spin torque efficiency and is odd under $\mathcal{T}$ symmetry. However, due to the two-point out-of-plane geometry required for both electrical STT writing and TMR-based electrical readout, STT encounters several limitations. A key challenge in STT-MRAMs is to maintain a sufficient separation between writing and readout currents to ensure robust signals while keeping writing currents below the tunnel barrier's breakdown threshold. In contrast, the transversal spin current induced by relativistic SOC effects such as spin Hall effect (SHE),[109] which is even under $\mathcal{T}$ symmetry, can decouple the reading and writing paths in MRAMs and thus enhance the device endurance.[110-113] Yet, it relies on relativistic SOC, which is typically much weaker than exchange coupling. Furthermore, the non-conserving nature of spin significantly restricts the spin diffusion length to merely nanoscale dimensions, severely limiting the practical utility of spin-orbit-torque magnetic random-access memories (SOT-MRAMs). AMs, with the anisotropic band splitting, offer a new perspective for the generation and manipulation of spin currents.[19, 90, 114-121]

As depicted in **Figure 6A**,[19] taking d-wave $RuO_2$ as an example, due to Fermi-surface anisotropies and the $[C_2||C_{4z}^{+}]$ symmetry, applying an electric field along the $[1\bar{1}0]$ axis results in parallel spin-up and spin-down charge currents that are different in magnitude. As a result, only longitudinal spin polarized current along the $[1\bar{1}0]$ axis is generated. In contrast, when the electric field is applied along the [100] axis, a transversal pure spin current is produced along the [010] axis. The spin polarization of this nonrelativistic spin current is aligned along the spin quantization axis of the electronic structure. Therefore, there are no symmetry requirements dictating that the applied field, current flow direction, and transported spin component be orthogonal to each other. The transversal spin current, generated by magnetic exchange splitting ($\mathcal{T}$-odd) and independent of SOC, can similarly engender spin splitter torque in spintronics, akin to the $\mathcal{T}$-even type. This phenomenon is termed “spin-splitter torque” (SST).[19] Characterized by high spin torque efficiency and a nonrelativistic origin, SST merges the advantages of STT and SOT while overcoming their respective key limitations. The calculated charge-spin conversion ratio for $RuO_2$ can reach up to 28% (Figure 6B), which is significantly larger than the most extensively exploited spin Hall angle of Pt.[109] Additionally, this ratio is only weakly impacted by SOC. Some literature refers to the phenomenon of generating such $\mathcal{T}$-odd spin currents, when considering SOC, as the magnetic spin Hall effect (MSHE).[122] To verify the aforementioned claims, experimentally, the ferromagnetic permalloy (Py) layer was deposited on the (100)- and (110)-oriented $RuO_2$ films (Figure 6C). When brought in contact with a ferromagnetic layer, the $RuO_2$(100) film generates a transversal spin current that induces both the SST and SOT, while for the (110)-oriented $RuO_2$ film, only SOT is exerted on the Py layer. The spin torque-ferromagnetic resonance (ST-FMR) measurements were performed to detect the spin current and spin torque generated by the $RuO_2$ films.[115] The ST-FMR spectra of $RuO_2$/Py samples reveal that $RuO_2$(100) film exerts much stronger spin torque on Py film (Figure 6D).[115] The calculated spin torque efficiency and spin torque conductivity in $RuO_2$ films further corroborate this finding , as illustrated in Figure 6E.[115] Furthermore, the measured spin Hall angle and spin diffusion length of $RuO_2$(100) film is 0.183 and greater than 12 nm, respectively, surpassing those of conventional spin source materials such as heavy metals.[123]

$RuO_2$ is not the only example of generating $\mathcal{T}$-odd spin currents. We point out that, unlike the anomalous Hall effect mentioned above, the existence of $\mathcal{T}$-odd spin currents is subject to fewer restrictions. All AMs, including both inorganic and organic types, are allowed to produce $\mathcal{T}$-odd transversal or longitudinal spin currents due to their broken $\mathcal{PT}$ and $\mathcal{T}\tau$ symmetries, but some require SOC as nonrelativistic SST effect is absent.[19] The $GdFeO_3$-type perovskite with C-type antiferromagnetic order were found to generate pure spin current under an applied electric field;[114] the spin-charge conversion ratio was expected to be over 30% in $V_2Te_2O$;[119] as the twist angle varies, the spin Hall angle of the twisted bilayer VOBr AM can reach a peak value of 1.4;[124] organic κ-Cl can act as a spin current generator by applying a thermal gradient, functioning as a time-reversal odd counterpart to the spin Nernst effect.[90] These theoretical and experimental findings, which demonstrate the observation of transversal spin current and spin splitting torque, suggest that AMs are not only ideal as information carriers free from Joule heating in electronic devices but also as promising spin source in spintronics.

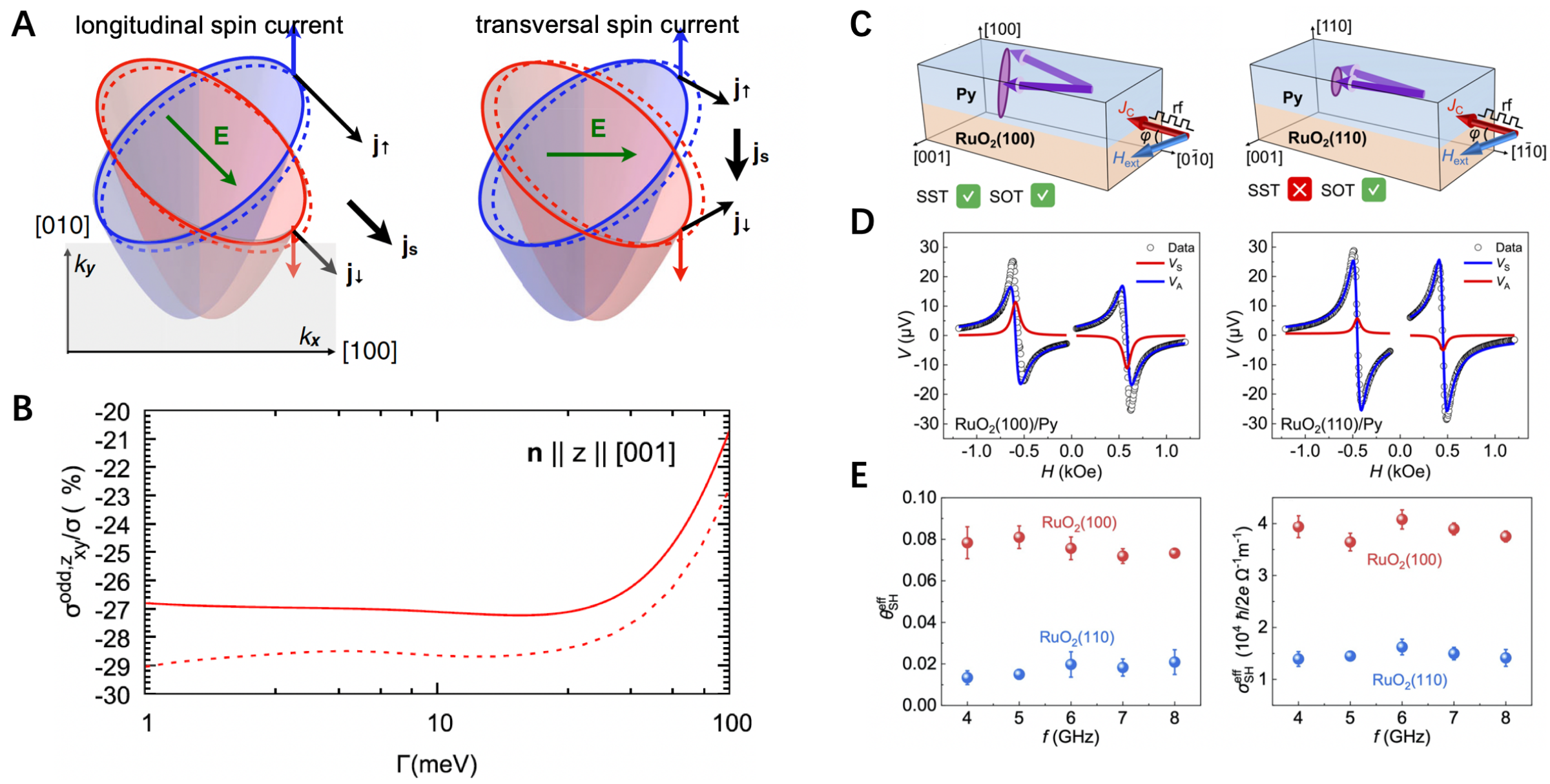

**Figure 6**. A) Schematic of the longitudinal spin current (left) and transversal spin current (right) in $RuO_2$. B) $\mathcal{T}$-odd charge-spin conversion ratio calculated with (solid line) and without (dashed line) SOC. The Néel vector is oriented along [001]. Reproduced with permission.[19] Copyright 2021, American Physical Society. C) Schematic of ST-FMR measurements for the $RuO_2(100)$/Py (left) and the $RuO_2(110)$/Py (right) samples. D) ST-FMR spectra of the $RuO_2(100)$/Py (left) and the $RuO_2(110)$/Py (right) samples. E) The calculated spin torque efficiency (left) and the spin torque conductivity (right) in $RuO_2(100)$ and $RuO_2(110)$ films. Reproduced with permission.[115] Copyright 2022, American Physical Society.

The spin splitting of altermagnetic bands and the associated spin currents proposed in $RuO_2$[19] make it possible to be used in GMR[21] and TMR[20-21] effects. GMR (TMR) effect refers to the phenomenon where the resistance of a magnetic tunnel junction—comprising a nonmagnetic metallic (insulating) layer sandwiched between two ferromagnetic electrodes—varies with the relative orientation of the magnetic order of the two ferromagnetic layers, and a higher spin polarization of the electrodes results in a large MR ratio.[125-127] A large magnitude of GMR ratio $(\rho_P - \rho_{AP})/(\rho_P + \rho_{AP})$, where P(AP) corresponds to (anti)parallel alignment of Néel vector in altermagnetic layers separated by nonmagnetic metallic spacer, up to 100% was predicted in magnetic tunnel junction comprised of $RuO_2$.[21] TMR in AMs has been studied both from first-principle and model calculations. [20-21] Shao *et al.* designed a $RuO_2/TiO_2/RuO_2$ tunnel junction (**Figure 7A**).[20] The $k_{\parallel}$-resolved transmission for spin-up and spin-down channel in P states clearly showed a large distribution of the spin-polarized conduction channels in $RuO_2$ (Figure 7B, left panel). In contrast, for the AP state, only the states near the zone center contributed to the transmission (Figure 7B, right panel), resulting in a significantly lower total transmission compared to the P state (Figure 7C). The calculated TMR ratio can reach up to 500% at the Fermi level (Figure 7D), a value comparable to that obtained in commercially used Fe/MgO/Fe (001) magnetic tunnel junction.[128-129] The superior performance suggests that altermagnetic TMR/GMR holds promise for commercial applications.

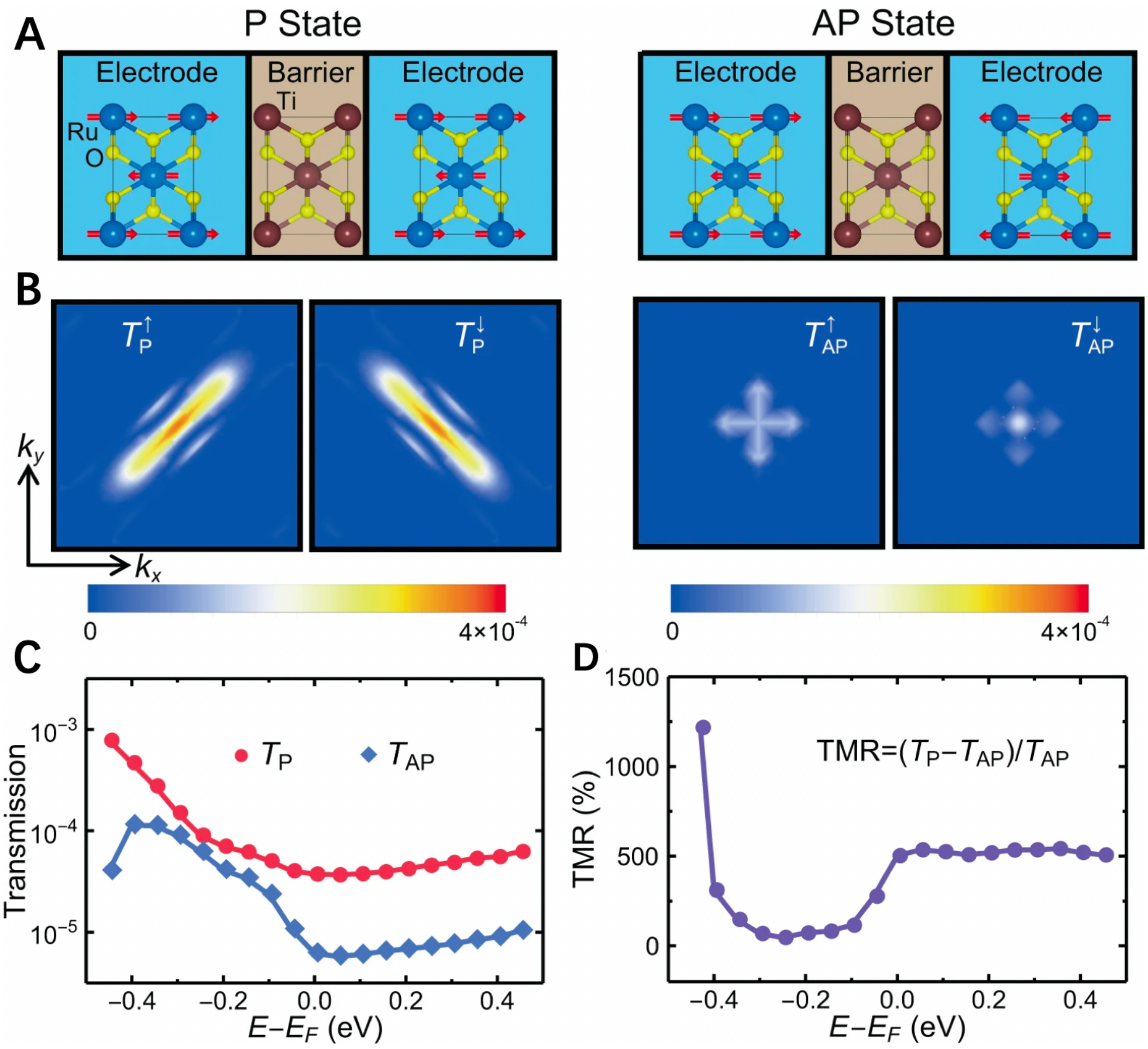

**Figure 7.** A) Structures of a TMR stack of $RuO_2/TiO_2/RuO_2$ tunnel junction with parallel (P) and antiparallel (AP) alignment of the Néel vectors. B) *Ab initio* calculated $k_{\parallel}$-resolved transmission of spin-up and spin down channel for P and AP states. C) Total transmissions with respect to energy for P and AP states. D) Energy dependence of calculated TMR. Reproduced under the terms of the CC-BY 4.0 license.[20] Copyright 2021, The Authors, Published by Springer Nature.

## 2.3. Magneto-optical effects

Magneto-optical (MO) effects, which represent the fundamental interplay between magnetism and light, play a pivotal role in condensed matter physics. Within the extensive array of MO effects, the Kerr and Faraday effects stand out, both of which are $\mathcal{T}$-odd MO effects. These phenomena are characterized by the rotation of the polarization plane of linearly polarized light as it reflects from or transmits through magnetic materials, respectively. The optical Hall conductivity, which is intimately related to the magneto-optical Kerr and Faraday effects (MOKE and MOFE), can be seen as the ac counterpart of anomalous Hall conductivity. Therefore, both MOKE and MOFE adhere to the same symmetry rules as the AHE, and consequently, these effects are also observed in AMs. In fact, as early as 1997, research predicted the presence of $\mathcal{T}$-odd MO effect in $LaMO_3$ (M=Cr, Mn, Fe), but it was not yet associated with the unconventional magnetic order.[130] Samanta *et al.* also identified crystal MO effect in the monolayer limit of an $SrRuO_3$ film, associating it with the lowering structural symmetry and antiferromagnetic order.[131] Zhou *et al.*,[132] employing first-principles calculations, identified MOKE and MOFE in AMs, which they designated as crystal chirality magneto-optical (CCMO) effects. These effects demonstrate that the signs of Kerr and Faraday spectra in $RuO_2$ and $CoNb_3S_6$ reverse with changes in material chirality, as illustrated in **Figures 8A-C**.[132] Moreover, the peak values of the Kerr and Faraday rotation angles induced by this novel CCMO effect—0.62 deg and $2.42 \times 10^5$ deg/cm for $RuO_2$, respectively—are significantly larger than those observed in traditional ferromagnetic and noncollinear antiferromagnetic materials,[133-135] suggesting their potential for applications in MO recording. In addition, a prominent MOKE has also been observed in altermagnetic $RuF_4$,[136] and the time-resolved MOKE has been employed to explore ultrafast spin dynamics in epitaxial MnTe(001)/InP(111) thin films.[137]

In addition to the MOKE and MOFE, x-ray magnetic circular dichroism (XMCD) also represents a highly significant $\mathcal{T}$-odd MO effect and also shares the same symmetry principles as AHE. Hariki *et al.* employed both *ab initio* theoretical models and experimental approaches to investigate XMCD in α-MnTe.[138] As shown in Figure 8D, the trends in XMCD, obtained both theoretically and experimentally, remained consistent regardless of the application of an external magnetic field, although the magnitudes differed significantly.[138] They attributed this phenomenon to the role of magnetic domains. In zero field, the cooling process causes the magnetic moments to cant, forming magnetic domains. Within each domain, there exists a ferrimagnetic-like alignment, with the direction of the magnetic moments varying between domains. Since XMCD is dependent on the direction of the magnetic moments, the effects from different domains cancel each other out under zero external magnetic field, resulting in a measured value that is ten times smaller than theoretical predictions. However, under a 6T external magnetic field, the magnetic moments within each domain align due to canting, thus closely aligning experimental and theoretical values (the experimentally induced magnetic moments were only $0.1\mu_B$ compared to $5\mu_B$ used in theoretical calculations, a difference of 50 times, yet the calculated results differed by only 80 times, nearly a 1:1 ratio). The conclusions are corroborated in Figure 8E, where an opposing external magnetic field leads to the reversal of ferrimagnetic-like moments within the domains, ultimately reversing the XMCD signal.[138] Hariki *et al.* computed XMCD at $L_{2,3}$ and $M_{2,3}$ edges of Ru in $RuO_2$ for different orientations of the Néel vector and the results for $M_{2,3}$ edges of Ru are shown in Figure 8F, indicating that the orientation of the altermagnetic order also has a significant effect on the MO effect. [139]

Moreover, a review systematically analyzed the nonlinear MO effects in antiferromagnetic and altermagnetic materials, revealing that altermagnetic domains with antiparallel Néel vector can be distinguished by the MO effects that are linearly proportional to the Néel vector.[140]

These findings indicate that MO tools are effective in probing altermagnetism, and MO effects can be utilized to investigate physical properties in AMs. Such insights are crucial for advancing research in magnetic materials and their technological applications.

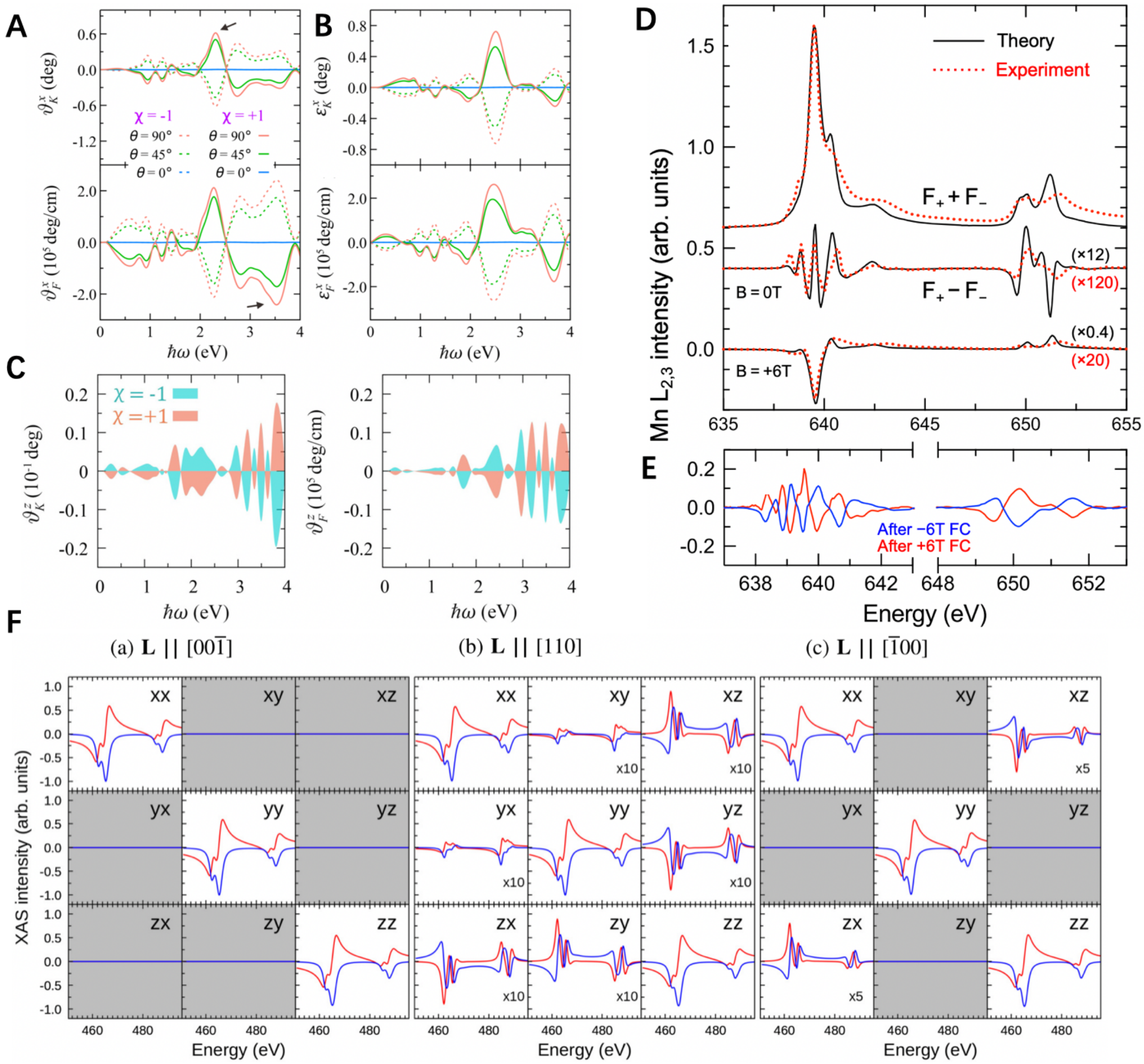

**Figure 8.** A) Kerr rotation and Faraday rotation angle and B) Kerr ellipticity and Faraday ellipticity for the left- and righthanded chirality states ($\chi = \pm 1$) of $RuO_2$. C) Kerr and Faraday rotation angles for the left- and right-handed chirality states of $CoNb_3S_6$. Reproduced with permission.[132] Copyright 2021, American Physical Society. D) Mn $L_{2,3}$-edge spectra, comparing theoretical results (solid black line) and experimental data (red dotted line), with light propagation along the c-axis. Top: X-ray absorption spectroscopy. Middle: XMCD in zero magnetic field following cooling in a 6 T field. Bottom: measured XMCD in an applied

magnetic field of 6 T, calculated XMCD with k aligned parallel to the magnetic moment, normalized by the ratio $m_{6T} = m_{sat} \approx 1/50$. E) XMCD following cooling in opposing magnetic fields. Reproduced with permission. [138] Copyright 2024, American Physical Society. F) Real (red) and imaginary (blue) components of the optical conductivity tensor at the Ru $M_{2,3}$ edge for various orientations of the Néel vector. Reproduced with permission.[139] Copyright 2024, American Physical Society.

### 2.4 Chiral magnons

Magnons, the spin wave quanta, are collective excitations of the spins observed in a magnetic system, that are central to the important research field called magnonics.[141-146] As depicted in **Figure 9A** , in the nonrelativistic limit and for collinear FMs, the magnons typically are polarized with one chirality (counterclockwise, cyan), have quadratic dispersion near the Γ point in the magnonic Brillouin zone, and are in GHz range. In contrast, for conventional collinear AFMs, the dispersions of magnons possess two chiral modes (cyan for counterclockwise and pink for clockwise) which are degenerate throughout the entire Brillouin zone and show a linear dispersion close to the Γ point.[74, 147-148] In addition, the resonance frequencies of antiferromagnetic magnons can reach up to THz range, making AFMs attractive for magnon spintronics when compared with FMs. [149-151] However, the net spin angular momentum is zero if these two opposite-chirality modes are equally populated. In order to observe and take advantages of the magnon-generated spin currents, the degeneracy between these two modes must be lifted by external stimuli (e.g., magnetic field).[148] Given that both FMs and AFMs have limitations in applying them in the realm of magnonics, it is natural to wonder whether there exists another type of magnetic matter that operated in the THz range and hosts nondegenerate magnonic chirality.

Recently, properties of magnons in AMs were theoretically investigated[152]. In these unconventional collinear compensated magnets, the degeneracy of magnon modes with opposite-chirality can be lifted without external stimuli and the sign of the splitting between opposite chirality magnons alternates throughout the Brillouin zone (Figure 9A).[152] Again, taking $RuO_2$ as an example, the nonrelativistic electronic and magnonic band structures along the high-symmetry paths at $k_z$=0 plane under zero magnetic field are shown in Figure 9B.[152] It is noted that both of these two kinds of splittings are strongly momentum dependent and the alternating sign of the chirality splitting is similar to that of the spin splitting along the same paths. The origin of the chirality splitting can be traced back to the anisotropic intra-sublattice exchange interactions, which must be accompanied by the spin group symmetry of altermagnetic phase. The anisotropic exchange alone is not sufficient condition for chiral magnons. This can be seen by analyzing an artificial antiferromagnetic phase of $RuO_2$, constructed by a 1 × 1 × 2 supercell. Such a supercell exhibits the anisotropic intra-sublattice exchange interactions but due to the antiferromagnetic spin group, the magnon bands are degenerate. Other AMs such as $CoF_2$, MnTe and $Fe_2O_3$ also showed alternating splitting (Figure 9C)[152] which means the alternating sign of the chirality splitting constitutes a distinctive hallmark of altermagnetic magnons with no counterpart in collinear FMs or AFMs. We notice that the magnitude of chiral splitting for $RuO_2$ was calculated to reach 10 meV, which is large enough to be detected experimentally via e.g. inelastic polarized neutron scattering.[153] The chiral splitting can also contribute to efficient spin Seebeck and spin Nernst effects of magnons.[154]

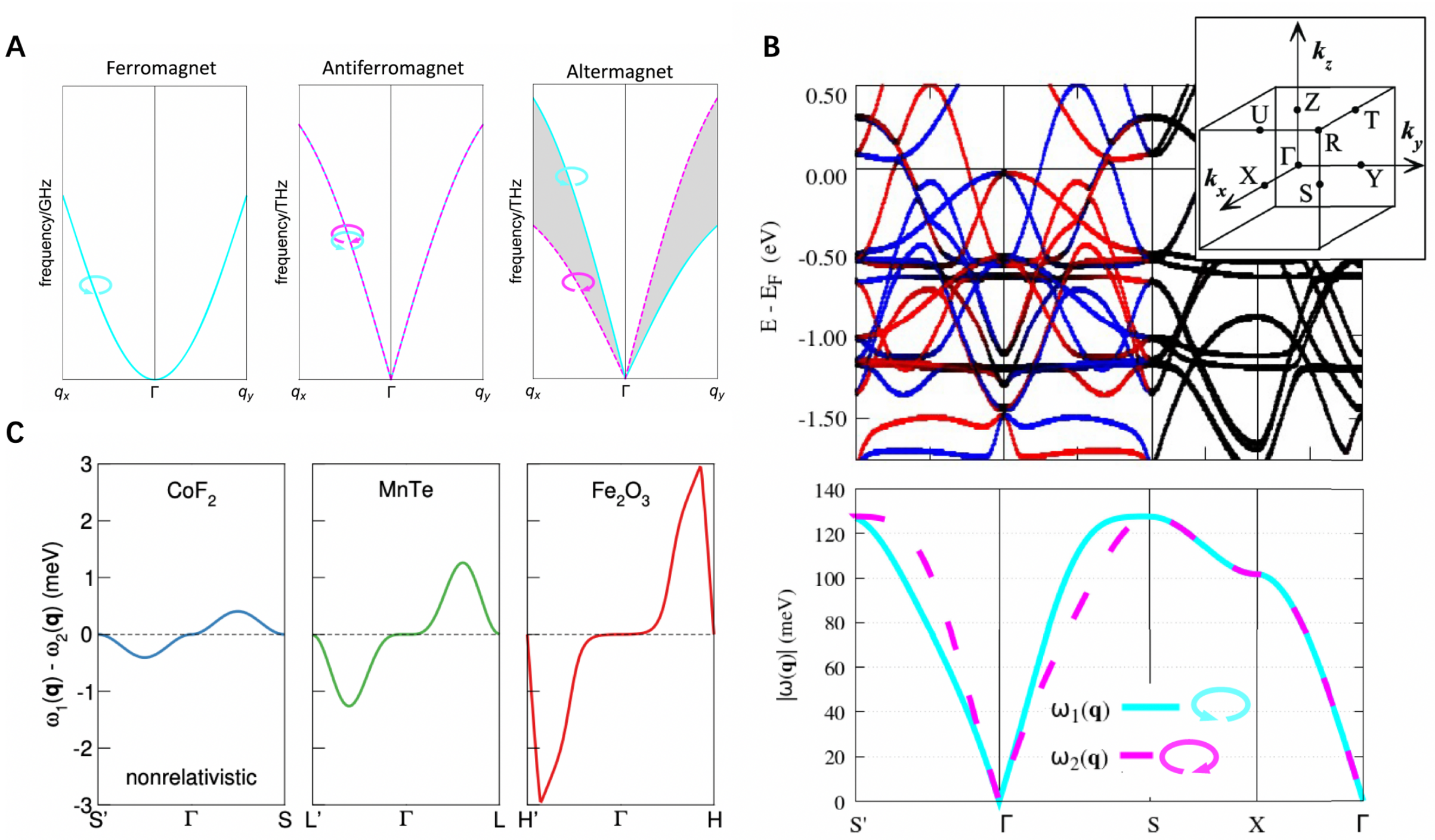

**Figure 9**. A) Schematic representation of the nonrelativistic magnon spectra in collinear FM, collinear AFM and AM. B) Up panel: the electronic band structure of the altermagnetic phase of $RuO_2$ with spin splitting. Down panel: corresponding magnonic band structure with alternating chirality splitting. C) The magnitude of chirality splitting in altermagnetic $CoF_2$, MnTe and $Fe_2O_3$. Reproduced with permission[152] Copyright 2023, American Physical Society.

## 3. Strategies to induce altermagnetism

Although there are many spontaneous altermagnetic candidates in nature, which we will mention in Section 4, there are always some situations in practice where it is necessary to manually construct AMs by functionalizing FMs or conventional AFMs. To achieve artificial AMs, the guiding principle is to break the combined $\mathcal{PT}$ and/or $\mathcal{T}\tau$ symmetry while preserving altermagnetic spin point group symmetries. Here, we briefly review four proposed routes towards altermagnetism.

### 3.1. Applied electric fields

In 3D magnetic materials, two spin-symmetry operations, $[C_2||\mathcal{P}][C_2||\mathcal{T}](\equiv \mathcal{PT})$ and $[C_2||\tau]$, enforce Kramers spin degeneracy throughout the Brillouin zone, while 2D systems have no out-of-plane components of momentum $k_\perp$, only in-plane components $k_\parallel$, therefore, violations of the aforementioned two symmetries are insufficient to realize spin splitting. For instance, $[C_2||C_{2z}][C_2||\mathcal{T}]$ symmetry transforms energy eigenstate $E(k_\parallel, s)$ as $[C_2||C_{2z}][C_2||\mathcal{T}]E(k_\parallel, s) = [C_2||C_{2z}]E(-k_\parallel, s) = E(k_\parallel, -s)$, resulting in spin degeneracy across the 2D Brillouin zone. $[C_2||\mathcal{M}_z]$ symmetry transforms energy eigenstate $E(k_\parallel, s)$ as $[C_2||\mathcal{M}_z]E(k_\parallel, s) = E(k_\parallel, -s)$, acting as $[C_2||\tau]$. The additional constraints imposed by symmetry considerations increase the difficulty of achieving 2D artificial AMs. Note that monolayer $MnPSe_3$ forms an antiferromagnetic arrangement, corresponding to the conventional collinear Néel order on the honeycomb lattice. The only symmetry that preserves spin degeneracy is $\mathcal{PT}$ symmetry. Consequently, monolayer $MnPSe_3$ is a good source for inducing altermagnetism.

Applied electric fields are one of the most common ways to tune and control physical properties.[155-163] As illustrated in **Figure 10A** ,[164] when an electric gate or a substrate is used to introduce an electric field perpendicular to the Mn plane in monolayer $MnPSe_3$, the two Se layers become crystallographically inequivalent due to the unequally additional electrostatic potential and thus the $\mathcal{PT}$ symmetry is broken, while the two opposite-spin Mn atoms are still locally equivalent and connected by glide-reflection symmetry. The resulting spin splitting is about 25 meV when applying an external electric field of 1 eV/Å. This mechanism can generate AHE in collinear AFMs but is different from layer-polarized anomalous Hall effects, in which the two magnetic layers become crystallographically inequivalent under the vertical electric field and are formally like ferrimagnets[161, 163]. Besides, we notice that a large spin splitting of 100 meV in the Fe bands can be seen in $FeSe/SrTiO_3$ heterostructure combined with quantum spin Hall effect, demonstrating the feasibility of combining nontrivial band-topology with unconventional altermagnetism in an experimentally known superconducting monolayer.[164]

### 3.2 Janus altermagnets

In addition to applied electric field, built-in electric field arising from a Janus structure can also break the inversion symmetry, thus enabling the transformation of a conventional AFM into an AM. For instance, in monolayer $MnPSe_3$, if one of the two Se layers is completely substituted with a different atomic species, such as S atoms (Figure 10B),[164] to form a Janus structure, the $\mathcal{PT}$ symmetry is broken while the two spin sublattices are mapped on top of each other by mirror rather than inversion or translational symmetries. The resulting band structure, in the absence of SOC, shows altermagnetic spin splitting of $i$-wave symmetry. Besides, the Janus structure could facilitate the integration of piezoelectricity, piezovalley, and piezomagnetism with altermagnetism, significantly expanding the versatility of AMs.[165-166]

### 3.3. Twisted bilayer altermagnets

"Twisting" has already been demonstrated, both theoretically and experimentally, to significantly change the physical properties of nonmagnetic and magnetic systems without alteration of chemical compositions.[167-173] Recently, twisted bilayer AMs have been proposed.[124, 174-175] In normally stacked antiferromagnetic bilayers, spin-up and spin-down states are degenerate in the band structure due to the high symmetry. Interestingly, introducing a twist between the two magnetic layers results in change of symmetry, which in turn causes significant k-dependent nonrelativistic spin splitting, introducing all the characteristic physical properties of altermagnetism.[174-175] Figure 10C illustrates the structure of twisted bilayer $MnPSe_3$ in which the magnetic order is intralayer and interlayer antiferromagnetic.[175] The special twist angle ($\boldsymbol{\theta = 21.79°}$) not only breaks $\boldsymbol{\mathcal{PT}}$ symmetry but also ensures the emergence of rotation symmetry ($[\boldsymbol{C_2} || \boldsymbol{C_{2[010]}}]$), which connects opposite spin sublattices. The bands exhibit spin degeneracy along high-symmetry paths, potentially leading to the misconception that it represents conventional antiferromagnetism. However, this is not applicable at any generic k-point, where no symmetry operations enforce the spin degeneracy. Therefore, a comprehensive analysis of the Brillouin Zone for AMs is essential. The maximum nonrelativistic spin splitting observed in twisted bilayer $MnPSe_3$ is 5.1 meV, which is relatively small compared to that of well-known bulk antiferromagnets but can be tuned by biaxial strain. Besides, other twist angles, such as 9.43° and 42.10°, can also induce altermagnetism. Therefore, twist angles represent a new degree of freedom for tuning the properties of AMs. It is worth mentioning that in twisted bilayer CrSBr, the transverse charge-to-spin conversion rate can reach up to 90%, demonstrating that AMs induced through this method hold significant potential for applications in spintronics.[174] We notice that the induction of spin splitting upon twisting a bilayer system with interlayer antiferromagnetic coupling is a universal phenomenon.

Liu *et al.* have proposed a more comprehensive analysis of twisted bilayer AMs and demonstrated that the spin-splitting character can range from $d$-wave to $i$-wave symmetry.[124]

### 3.4. Supercell altermagnets

The above three strategies are applicable in the case of 2D AMs. Here, a method that can induce 3D altermagnetism is proposed. Until now, the exploration of new altermagnetic candidates in nature has predominantly concentrated on materials where the magnetic unit cell coincides with the nonmagnetic one. A very recent work has expanded the family of altermagnetic candidates by enlarging the magnetic unit cell.[176] Figure 10D shows two magnetic structures of $MnSe_2$ (termed $MnSe_2$-I and $MnSe_2$-II) and their band structure.[176] Both structures are constructed by stacking three nonmagnetic unit cells along the c-axis. The distribution of the magnetic moments varies between them, which is revealed by the different colored spin density around Mn atoms. The anisotropic spin density also indicates that in both cases, two sites with opposite spin cannot be linked by a straightforward antiunitary translation. The glide planes connecting Mn with opposite spin for $MnSe_2$-I and $MnSe_2$-II are $[\mathcal{C}_2||\,\mathcal{M}_y]$ and $[\mathcal{C}_2||\,\mathcal{M}_z]$ (disregarding translational symmetry for simplicity), respectively. Due to the differing glide planes, spin splitting occurs along distinct k-path, with the maximum values of spin splitting being 167 meV and 71 meV, respectively. Symmetry analysis has revealed that both $MnSe_2$-I and $MnSe_2$-II are d-wave AMs. Notably, the energy difference of $MnSe_2$-I and $MnSe_2$-II is merely 5 meV. This minimal discrepancy suggests that changing the ground state through an external perturbation could be feasible. Despite the current lack of experimental techniques for manipulating the magnetic order of individual atoms, the proposal of supercell AMs encourages further exploration in the design of AMs.

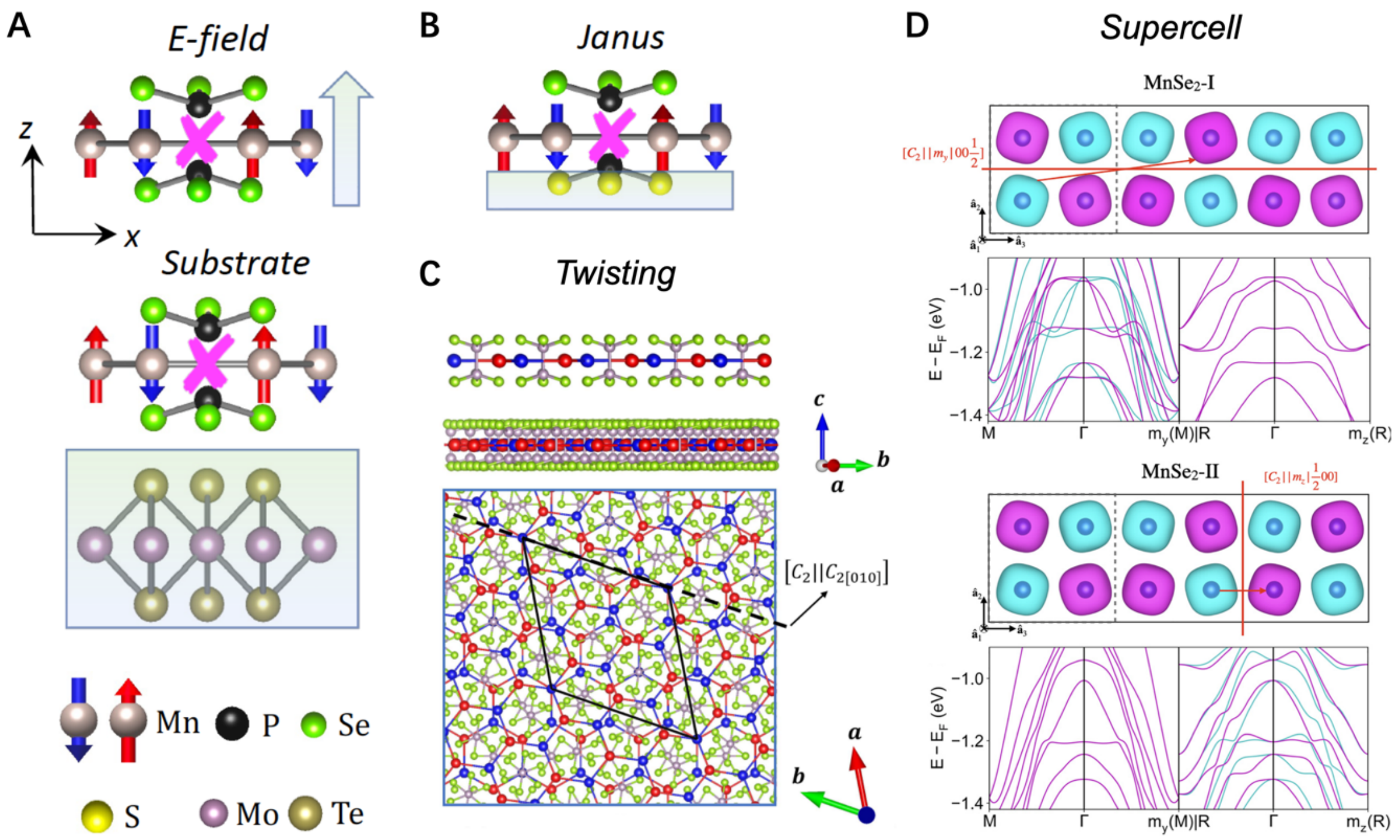

**Figure 10**. Four mechanisms of inducing altermagnetic phase. A) Top: applying external electric field. Bottom: placing the monolayer on a substrate. B) Forming a Janus structure by replacing the bottom Se atoms with S atoms. Reproduced with permission.[164] Copyright 2024, The Authors, Published by arXiv. C) Twisting bilayer of $MnPSe_3$. Reproduced with permission.[175] Copyright 2024, American Physical Society. D) Supercell magnetic configurations of $MnSe_2$-I (up) and $MnSe_2$-II (down) and their corresponding nonrelativistic band structures. Cyan and magenta denote positive and negative values of the spin density, respectively. Reproduced with permission.[176] Copyright 2024, American Physical Society.

## 4. Altermagnetic candidates

In this Section, we review the altermagnetic candidates, neglecting the artificial AMs mentioned in Section 3. Before moving on to this Section, we recall the criteria necessary to identify altermagnetic phase[26]: an even number of magnetic atoms in the magnetic unit cell; no $\mathcal{PT}$ and $\mathcal{T}\tau$ symmetries; the sign of the spin splitting alternates across the Brillouin zone; the two opposite-spin sublattices are linked by a real space rotation transformation (proper or improper, and symmorphic or non-symmorphic). It is worth noting that symmetry constraints prevent the emergence of one-dimensional (1D) AMs, owing to the absence of rotational transformations in reciprocal space for 1D systems. Therefore, we focus exclusively on 2D and 3D altermagnetic candidates, encompassing a diverse spectrum of conduction types ranging from insulators and metals to superconductors. In addition, a program was recently developed to determine the magnetic phase of symmetry-compensated collinear magnetic material. We refer interested readers to a detailed information in ref.[177].

### 4.1 2D altermagnetic candidates

As discussed in Section 3.1, the symmetry requirements for 2D AMs are relatively more stringent compared to their bulk counterparts. Consequently, there are fewer viable candidates for 2D AMs than for 3D ones. Nevertheless, considerable efforts have been made to search for and study the properties of 2D AMs. For instance, Sødequist and Olsen [178] identified seven 2D AMs using high-throughput calculations based on the Computational 2D Materials Database (C2DB), while Liu *et al.*[179] computationally derived two 2D AMs from the 2D NiAs-type non-van der Waals family, providing new perspectives for the synthesis of 2D AMs. In addition, Zeng and Zhao explored the 2D spin layer group, which is conducive to the further understanding of 2D AMs.[180] These materials are listed in **Table 1**, which includes those currently considered to be 2D AMs. It is anticipated that ongoing research will further expand the family of 2D altermagnetic candidates.

**Table 1.** 2D altermagnetic candidates with their corresponding nonmagnetic space group (NSG), magnetic space group (MSG), and even-parity spin symmetry type. The number of NSG represents in parentheses.

| No. | Compound | NSG | MSG | Spin symmetry | Ref. |
|---|---|---|---|---|---|
| 1 | $OsNNaSCl_5$ | $P2_1$ (4) | $P2_1$ | *d*-wave | [178] |
| 2 | $V_2ClBrI_2O_2$ | *Cm* (8) | *P1* | *d*-wave | [178] |
| 3 | $RuF_4$ | $P2_1/c$ (14) | $P2_1/c$ | *d*-wave | [88] |
| 4 | $VF_4$ | $P2_1/c$ (14) | $P2_1/c$ | *d*-wave | [178] |
| 5 | $AgF_2$ | $P2_1/c$ (14) | $P2_1'/c'$ | *d*-wave | [178] |
| 6 | $OsF_4$ | $P2_1/c$ (14) | $P2_1/c$ | *d*-wave | [178] |
| 7 | $SrRuO_3$(010) | $P2_1/c$ (14) | $P2_1/c$ | *d*-wave | [131] |
| 8 | $MnTeMoO_6$ | $P2_12_12$ (18) | | *d*-wave | [180] |
| 9 | FeS(110) | *Pmma* (51) | | *d*-wave | [179] |
| 10 | FeSe(110) | *Pmma* (51) | | *d*-wave | [179] |
| 11 | $VP_2H_8(NO_4)_2$ | *P4bm* (100) | | *g*-wave | [180] |
| 12 | $Fe_2WTe_4$ | $P\bar{4}2m$ (111) | | *d*-wave | [181] |
| 13 | $Fe_2MoZ_4$ (Z=S,Se,Te) | $P\bar{4}2m$ (111) | | *d*-wave | [181] |
| 14 | $Ca(CoN)_2$ | $P\bar{4}m2$ (115) | $P\bar{4}'m'2$ | *d*-wave | [182] |
| 15 | $Ca(FeN)_2$ | $P\bar{4}m2$ (115) | $P\bar{4}'m'2$ | *d*-wave | [183] |
| 16 | CrO | *P4/mmm* (123) | *P4'/mmm'* | *d*-wave | [184-185] |
| 17 | $V_2Se_2O$ | *P4/mmm* (123) | *Pm'm'm* | *d*-wave | [17] |
| 18 | $V_2Te_2O$ | *P4/mmm* (123) | *P4'/mm'm* | *d*-wave | [119, 186] |
| 19 | $Cr_2Te_2O$ | *P4/mmm* (123) | | *d*-wave | [154] |
| 20 | $Cr_2Se_2O$ | *P4/mmm* (123) | | *d*-wave | [154] |
| 21 | $Fe_2Se_2O$ | *P4/mmm* (123) | *P4'/mm'm* | *d*-wave | [187] |
| 22 | FeSe(001) | *P4/nmm* (129) | | *d*-wave | [164, 188] |
| 23 | 2H-$FeBr_3$ | $P\bar{6}2m$ (189) | $P\bar{6}2m$ | *i*-wave | [178] |

### 4.2. 3D altermagnetic candidates

After reviewing the literature on 3D altermagnetic candidates, we propose that perovskite materials are particularly favorable for altermagnetism. This includes those with $GdFeO_3$-type and $LiNbO_3$-type structures, as well as perovskites that are structurally doubled or tripled, in which the distortion of the O octahedra reduces the symmetry of the system, making the emergence of altermagnetism possible. Additionally, 3D rutile fluorides or oxides along with other fluorides and sulfate fluorides, have also emerged as promising candidates. Moreover, altermagnetism is not limited to inorganic materials; it can also occur in organic materials.

We tabulate 221 3D altermagnetic candidates in **Table 2**, mainly referring to the work of Šmejkal *et al.*,[9, 24, 26] Guo *et al.*,[89] Chen *et al.*[189] and Gao *et al.*[190] Šmejkal and colleagues were pioneers in applying spin group theory to identify several important AMs, including MnTe, CrSb, $Mn_5Si_3$, $RuO_2$, Co/$VNb_3S_6$, and $Fe_2O_3$. [9, 14, 24, 26] Among these, MnTe, CrSb, $Mn_5Si_3$, and $RuO_2$ have been experimentally confirmed as AMs.[14, 31, 33-35, 64, 67-71, 93, 95, 97, 101, 104, 116-118] Guo *et al.* [89] and Gao *et al.* [190] further selected altermagnetic candidates through a large-scale *ab initio* study and AI search engine, respectively. Chen *et al.* systematically developed the Spin Space Group theory to describe collinear magnetic configurations, identified 422 altermagnetic spin space groups, and identified potential altermagnetic candidates using the MAGNDATA database.[189]

**Table 2.** 3D altermagnetic candidates with their corresponding NSG, MSG, and Néel temperature. The number of NSG represents in parentheses. The experimentally indicated AMs via measurements of spin-splitting spectrum and anomalous transport are highlighted in bold.

| No. | Compound | NSG | MSG | Néel temperature [K] | Ref. |
|---|---|---|---|---|---|
| 1 | **$KV_2Se_2O$** | *P4/mmm* (123) | *P4'/mm'm* | ≥ RT | [191] |
| 2 | **$Rb_{1-\delta}V_2Te_2O$** | *P4/mmm* (123) | *P4'/mm'm* | ≥ 307 | [192] |
| 3 | **$RuO_2$** | $P4_2/mnm$ (136) | $P4_2'/mnm'$ | > 300 | [9-10, 17, 72, 84, 95, 101, 115-118, 123] |
| 4 | **$Mn_5Si_3$** | *$P6_3/mcm$* (193) | $P\bar{1}$ | 240 | [14, 33-34, 102, 104] |
| 5 | **CrSb** | *$P6_3/mmc$* (194) | *$P6_3'/m'm'c$* | > 600 | [24, 35, 67-71, 193] |
| 6 | **MnTe** | *$P6_3/mmc$* (194) | *Cm'c'm* | 310 | [17, 24, 31, 64-65, 97, 137-138, 194-195] |
| 7 | $LiFeP_2O_7$ | *$P2_1$* (4) | *$P2_1$* | 22 | [9, 196] |
| 8 | $CaLaFeAgO_6$ | *Pc* (7) | | | [190] |
| 9 | $Mn_4Nb_2O_9$ | *Cc* (9) | *Cc* | 52.1(1) | [189, 197] |
| 10 | $CaFe_5O_7$ | *$P2_1/m$* (11) | *$P2_1'/m'$* | 125 | [189, 198] |
| 11 | $Fe_3F_8(H_2O)_2$ | *C2/m* (12) | *C2'/m'* | 157 | [189, 199] |
| 12 | $RbMnF_4$ | *$P2_1/a$* (14) | $P\bar{1}$ | 3.7(1) | [89, 200] |
| 13 | $Li_2Co(SO_4)_2$ | *$P2_1/c$* (14) | *$P2_1'/c'$* | 7~8 | [89, 201] |
| 14 | $LiFe(SO_4)_2$ | *$P2_1/c$* (14) | *$P2_1/c$* | 35~39 | [201] |
| 15 | $Li_2Mn(SO_4)_2$ | *$P2_1/c$* (14) | *$P2_1/c$* | 6 | [201] |
| 16 | $Fe_3(PO_4)_2(OH)_2$ | *$P2_1/c$* (14) | *$P2_1/c$* | 160 | [89, 202] |
| 17 | $CuF_2$ | *$P2_1/c$* (14) | | 69 | [24, 203] |
| 18 | $Li_2Ni(SO_4)_2$ | *$P2_1/c$* (14) | *$P2_1/c$* | 15 | [189, 204] |
| 19 | $K_2ReI_6$ | *$P2_1/n$* (14) | *$P2_1/c$* | 24 | [89, 205] |
| 20 | $Sr_2CoTeO_6$ | *$P2_1/n$* (14) | *$P2_1/c$* | 15~19 | [189, 206] |
| 21 | $Sr_2ScOsO_6$ | *$P2_1/n$* (14) | *$P2_1/c$* | 92 | [189, 207] |
| 22 | $Sr_2YRuO_6$ | *$P2_1/n$* (14) | *$P2_1/c$* | 26 | [189, 208] |
| 23 | $Sr_2TbIrO_6$ | *$P2_1/n$* (14) | *$P2_1/c$* | 51 | [189, 209] |
| 24 | $Sr_2LuRuO_6$ | *$P2_1/n$* (14) | *$P2_1/c$* | 30(1) | [189, 210] |
| 25 | $Sr_2YbRuO_6$ | *$P2_1/n$* (14) | *$P2_1/c$* | 36 | [189, 211] |
| 26 | $Sr_2TmRuO_6$ | *$P2_1/n$* (14) | *$P2_1/c$* | 36 | [189, 212] |
| 27 | $Sr_2TbRuO_6$ | *$P2_1/n$* (14) | *$P2_1/c$* | 41 | [189, 213] |
| 28 | $Sr_2HoRuO_6$ | *$P2_1/n$* (14) | *$P2_1/c$* | 36 | [189, 214] |
| 29 | $Sr_2DyRuO_6$ | *$P2_1/n$* (14) | *$P2_1/c$* | 39.5 | [189, 215] |
| 30 | $Mn_2ScSbO_6$ | *$P2_1/n$* (14) | *$P2_1/c$* | 22.3 | [189, 216] |
| 31 | $CuAg(SO_4)_2$ | *$P2_1/n$* (14) | | | [217-218] |
| 32 | $La_2LiRuO_6$ | *$P2_1/n$* (14) | *$P2_1/c$* | 30 | [189, 219] |
| 33 | $Sr_2Co_{0.9}Mg_{0.1}TeO_6$ | *$P2_1/n$* (14) | *$P2_1/c$* | 13 | [189, 206] |
| 34 | $Sr_2CoOsO_6$ | *B2/c* (15) | *C2/c* | 108 | [89, 220] |
| 35 | $Ba_3CoIr_2O_9$ | *C2/c* (15) | *C2/c* | 107 | [89, 221] |
| 36 | $FeSO_4F$ | *C2/c* (15) | *C2'/c'* | 100 | [89, 222] |
| 37 | $FeOHSO_4$ | *C2/c* (15) | *C2'/c'* | 125 | [89, 223] |
| 38 | $BiCrO_3$ | *C2/c* (15) | *C2/c* | | [9] |
| 39 | $BaCrF_5$ | *$P2_12_12_1$* (19) | *$P2_1'2_1'2_1$* | 3.4 | [89, 224] |
| 40 | $FeHO_2$ | *$Pmn2_1$* (31) | | > RT | [190, 225-226] |
| 41 | $CaLaCr_2O_6$ | *$Pmn2_1$* (31) | | | [190] |
| 42 | $Y_2Cu_2O_5$ | *Pna2₁* (33) | *Pna2₁* | 13 | [89, 227-228] |
| 43 | $NaFeO_2$ | *Pna2₁* (33) | | | [190, 229] |
| 44 | $[C(ND_2)_3]Cu(DCOO)_3$ | *Pna2₁* (33) | *Pn'a'2₁* | | [189, 230] |
| 45 | $[C(ND_2)_3]Cu(DCOO)_3$ | *Pna2₁* (33) | *Pna2₁* | | [189, 230] |
| 46 | $Ca_3Mn_2O_7$ | *Cmc2₁* (36) | *Cm'c2₁'* | 115 | [89, 231] |
| 47 | $Ca_3Cr_2O_7$ | *Cmc2₁* (36) | | | [190] |
| 48 | $ErGe_{1.83}$ | *Cmc2₁* (36) | *Cmc2₁* | 6 | [189, 232] |
| 49 | $Cu_2V_2O_7$ | *Fdd2* (43) | *Fd'd'2* | 33.4(1) | [89, 233] |
| 50 | $ZnFeF_5(H_2O)_2$ | *Imm2* (44) | *Imm2* | 9(2) | [89, 234] |
| 51 | $Sr_2MnGaO_5$ | *Ima2* (46) | *Im'a2'* | 180 | [189, 235] |
| 52 | $BiFe_{0.5}Sc_{0.5}O_3$ | *Ima2* (46) | *Im'a2'* | ~ 220 | [189, 236] |
| 53 | $[C(ND_2)_3]Co(DCOO)_3$ | *Pnna* (52) | *Pn'na'* | 14.04(1) | [189, 230] |
| 54 | $[C(ND_2)_3]Mn(DCOO)_3$ | *Pnna* (52) | *Pn'n'a* | 8.73(2) | [189, 230] |
| 55 | $Fe_{1.5}Mn_{1.5}BO_5$ | *Pbam* (55) | *Pbam* | 100 | [189, 237] |
| 56 | $Mn(N(CN)_2)_2$ | *Pnnm* (58) | *Pnn'm'* | 16 | [89, 238] |
| 57 | $FeSb_2$ | *Pnnm* (58) | *Pnn'm'* | | [15] |
| 58 | $CrSb_2$ | *Pnnm* (58) | | 273 | [15, 239-240] |
| 59 | $Eu_3In_2As_4$ | *Pnnm* (58) | | | [241-242] |
| 60 | $Fe_2WO_6$ | *Pbcn* (60) | *Pbc'n'* | | [189, 243] |
| 61 | $Ca_2RuO_4$ | *Pbca* (61) | | | [244] |
| 62 | $CaCrO_3$ | *Pbnm* (62) | *Pb'n'm* | 90 | [24, 245-246] |
| 63 | $YCrO_3$ | *Pbnm* (62) | *Pn'ma'* | 141 | [89, 247] |
| 64 | $TbCrO_3$ | *Pbnm* (62) | *Pn'm'a* | 158 | [89, 247-248] |
| 65 | $DyCrO_3$ | *Pbnm* (62) | *Pn'm'a* | | [189, 249] |
| 66 | $ErCrO_3$ | *Pbnm* (62) | *Pn'ma'* | 133 | [189, 247] |
| 67 | $CeFeO_3$ | *Pbnm* (62) | *Pnma* | 220 | [189, 250] |
| 68 | $CeFeO_3$ | *Pbnm* (62) | *Pn'ma'* | 720 | [189, 250] |
| 69 | $SmFeO_3$ | *Pbmn* (62) | *Pn'm'a* | 480 | [89, 251] |
| 70 | $SmFeO_3$ | *Pnma* (62) | *Pn'ma'* | 675 | [89, 251] |
| 71 | $TbFeO_3$ | *Pbnm* (62) | *Pn'ma'* | 681 | [17, 89, 252] |
| 72 | $DyFeO_3$ | *Pbnm* (62) | *Pnma* | 73 | [189, 253] |
| 73 | $DyFeO_3$ | *Pbnm* (62) | *Pn'ma'* | | [189, 253] |
| 74 | $ErVO_3$ | *Pbnm* (62) | *P2₁/c* | 56 | [189, 254] |
| 75 | $NdCoO_3$ | *Pbnm* (62) | *Pnma* | 1.20(1) | [189, 255] |
| 76 | $TbCr_{0.5}Mn_{0.5}O_3$ | *Pbnm* (62) | *Pn'ma'* | ~ 84 | [189, 256] |
| 77 | κ-Cl | *Pnma* (62) | *Pn'ma'* | 23 | [26, 90, 257] |
| 78 | $YVO_3$ | *Pnma* (62) | *Pn'm'a* | 77 | [189, 258] |
| 79 | $NdVO_3$ | *Pnma* (62) | *P2₁'/m'* | 135 | [189, 258] |
| 80 | $LaCrO_3$ | *Pnma* (62) | *Pnma* | 289 | [89, 259] |
| 81 | $LaCrO_3$ | *Pnma* (62) | *Pn'ma'* | | [17, 89, 260] |
| 82 | $ScCrO_3$ | *Pnma* (62) | *Pnma* | 73 | [89, 261] |
| 83 | $InCrO_3$ | *Pnma* (62) | *Pnma* | 93 | [89, 261] |
| 84 | $TlCrO_3$ | *Pnma* (62) | *Pnma* | 89 | [89, 261] |
| 85 | $ZrCrO_3$ | *Pnma* (62) | | | [190] |
| 86 | $YRuO_3$ | *Pnma* (62) | *Pn'ma'* | 97 | [189, 262] |
| 87 | $LaMnO_3$ | *Pnma* (62) | *Pn'ma'* | 139.5 | [9, 17, 263] |
| 88 | $CaMnO_3$ | *Pnma* (62) | | 120 | [9, 264] |
| 89 | $Ca_{1-x}Ce_xMnO_3$ | *Pnma* (62) | | | [9] |
| 90 | $NdMnO_3$ | *Pnma* (62) | *P2₁/m* | 85 | [189, 265] |
| 91 | $PrMnO_3$ | *Pnma* (62) | *Pn'ma'* | 91 | [189, 265] |
| 92 | $LaFeO_3$ | *Pnma* (62) | *Pn'ma'* | 750 | [266] |
| 93 | $NdFeO_3$ | *Pnma* (62) | *Pn'ma'* | 760 | [89, 267] |
| 94 | $NaOsO_3$ | *Pnma* (62) | *Pn'ma'* | 411.(2) | [17, 89, 268] |
| 95 | $Fe_2PO_5$ | *Pnma* (62) | *Pnma* | 250 | [89, 269-270] |
| 96 | $NiFePO_5$ | *Pnma* (62) | *Pnma* | 178(5) | [89, 271] |
| 97 | $CuFePO_5$ | *Pnma* (62) | *Pnma* | 195(5) | [89, 271] |
| 98 | $Mn_2SeO_3F_2$ | *Pnma* (62) | *Pn'ma'* | 26 | [89, 272] |
| 99 | $MnO_2$ | *Pnma* (62) | | | [190, 273] |
| 100 | $KMnF_3$ | *Pnma* (62) | *Pn'ma'* | 75 | [24, 274] |
| 101 | $Ca_2MnGaO_5$ | *Pnma* (62) | *Pnm'a'* | 160 | [189, 235] |
| 102 | $Ca_2PrCr_2TaO_9$ | *Pnma* (62) | *Pn'm'a* | 130 | [189, 275] |
| 103 | $Ca_2PrCr_2NbO_9$ | *Pnma* (62) | *Pn'm'a* | 110 | [189, 275] |
| 104 | $Ho_{0.2}Bi_{0.8}FeO_3$ | *Pnma* (62) | *Pn'ma'* | | [189, 276] |
| 105 | $Ho_{0.15}Bi_{0.85}FeO_3$ | *Pnma* (62) | *Pn'ma'* | | [189, 276] |
| 106 | $La_{0.95}Ba_{0.05}Mn_{0.95}Ti_{0.05}O_3$ | *Pnma* (62) | *Pn'ma'* | | [189, 277] |
| 107 | $La_{0.75}Bi_{0.25}Fe_{0.5}Cr_{0.5}O_3$ | *Pnma* (62) | *Pnma* | 350 | [189, 278] |
| 108 | $La_{0.5}Sr_{0.5}FeO_{2.5}F_{0.5}$ | *Pnma* (62) | *Pn'ma'* | | [189, 279] |
| 109 | $BiFe_{0.5}Sc_{0.5}O_3$ | *Pnma* (62) | *Pn'm'a* | ~ 220 | [189, 236] |
| 110 | $Bi_{0.85}Ca_{0.15}Fe_{0.55}Mn_{0.45}O_3$ | *Pnma* (62) | *Pn'm'a* | | [189, 280] |
| 111 | $Ca_2Fe_{0.875}Cr_{0.125}GaO_5$ | *Pnma* (62) | *Pn'm'a* | | [189, 281] |
| 112 | $Pb_2Mn_{0.6}Co_{0.4}WO_6$ | *Pmcn* (62) | *Pm'c21'* | 9 | [189, 282] |
| 113 | $NiCrO_4$ | *Cmcm* (63) | *Cmcm* | | [189, 283] |
| 114 | $Tb_2Ir_3Ga_9$ | *Cmcm* (63) | *Cm'cm'* | 12.5 | [189, 284] |
| 115 | $CaIrO_3$ | *Cmcm* (63) | *Cm'cm'* | 115 | [89, 285-286] |
| 116 | $CoFe_3O_5$ | *Cmcm* (63) | *Cm'cm'* | ~ 300 | [189, 287] |
| 117 | $Nd_2PdGe_6$ | *Cmce* (64) | *Cm'c'a* | 4 | [189, 288] |
| 118 | $LaCaFeO_4$ | *Cmce* (64) | *Cm'c'a* | | [189, 289] |
| 119 | $Sr_4Fe_4O_{11}$ | *Cmmm* (65) | *Cmm'm'* | 232(4) | [89, 290] |
| 120 | $La_2NiO_3F_2$ | *Cccm* (66) | | 49 | [291-292] |
| 121 | $YBaMn_2O_{5.5}$ | *Ibam* (72) | *C2/m* | ~ 140 | [189, 293] |
| 122 | $YBaMn_2O_{5.5}$ | *Ibam* (72) | *Ib'a'm* | ~ 120 | [189, 293] |
| 123 | $Y_2SrCu_{0.6}Co_{1.4}O_{6.5}$ | *Ibam* (72) | *Ib'a'm* | 385 | [189, 294] |
| 124 | $Bi_{0.8}La_{0.2}Fe_{0.5}Mn_{0.5}O_3$ | *Imma* (74) | *Imm'a'* | ~ 240 | [189, 295] |
| 125 | α-$MnO_2$ | *I4/m* (87) | | | [23, 296] |
| 126 | $KRu_4O_8$ | *I4/m* (87) | | | [24] |
| 127 | $K_yFe_{2-x}Se_2$ | *I4/m* (87) | *C2'/m'* | | [189, 297] |
| 128 | $Rb_yFe_{2-x}Se_2$ | *I4/m* (87) | *C2'/m'* | | [189, 297] |
| 129 | NaFeO2 | $P4_12_12$ (92) | | | [190, 298] |
| 130 | GdAlSi | $I4_1md$ (109) | $I4_1'm'd$ | 32 | [299] |
| 131 | $SrMn_2V_2O_8$ | $I4_1cd$ (110) | *Ib'a2'* | 42.2 | [89, 300] |
| 132 | $Ba_2MnSi_2O_7$ | $P\bar{4}2_1m$ (113) | $P\bar{4}2_1m$ | 2.95 | [89, 301] |
| 133 | $Ba_2CoGe_2O_7$ | $P\bar{4}2_1m$ (113) | *Cm'm2'* | 6.7 | [89, 302] |
| 134 | $Ca_2CoSi_2O_7$ | $P\bar{4}2_1m$ (113) | $P2_12_1'2'$ | 5.7 | [189, 303] |
| 135 | $CsCoF_4$ | $I\bar{4}C2$ (120) | $I\bar{4}'$ | 54 | [189, 304] |
| 136 | $CuFeS_2$ | $I\bar{4}2d$ (122) | $I\bar{4}2d$ | > RT | [189, 305] |
| 137 | $UCr_2Si_2C$ | *P4/mmm* (123) | *Pm'm'm* | | [189, 306] |
| 138 | $ZrMn_2Ge_4O_{12}$ | *P4/nbm* (125) | *P4'/nbm'* | 8 | [89, 307] |
| 139 | $CeMn_2Ge_4O_{12}$ | *P4/nbm* (125) | *P4'/nbm'* | 7.6 | [189, 308] |
| 140 | $CeMnCoGe_4O_{12}$ | *P4/nbm* (125) | *Pb'an'* | 5.8 | [189, 308] |
| 141 | $Nb_2FeB_2$ | *P4/mbm* (127) | *P4/mb'm'* | | [99] |
| 142 | $Ta_2FeB_2$ | *P4/mbm* (127) | *P4/mb'm'* | | [99] |
| 143 | $NdB_2C_2$ | *P4/mbm* (127) | | 8.8 | [190, 309] |
| 144 | $Mg_2FeIr_5B_2$ | *P4/mbm* (127) | | | [190, 310] |
| 145 | $Mg_2MnIr_5B_2$ | *P4/mbm* (127) | | | [190, 310] |
| 146 | $Mg_2NiIr_5B_2$ | *P4/mbm* (127) | | | [190, 310] |
| 147 | $Sc_2MnIr_5B_2$ | *P4/mbm* (127) | | | [190, 311] |
| 148 | $MnF_2$ | $P4_2/mnm$ (136) | $P4_2'/mnm'$ | 67.4 | [17, 24, 312] |
| 149 | $CoF_2$ | $P4_2/mnm$ (136) | $P4_2'/mnm'$ | 37.7 | [17, 24, 313] |
| 150 | $NiF_2$ | $P4_2/mnm$ (136) | *Pnn'm'* | 73.2 | [17, 189, 314] |
| 151 | $ReO_2$ | $P4_2/mnm$ (136) | $P4_2'/mnm'$ | | [315] |
| 152 | β-$MnO_2$ | $P4_2/mnm$ (136) | | | [23-24] |
| 153 | $LiFe_2F_6$ | $P4_2/mnm$ (136) | $P4_2'/mnm'$ | 105 | [89, 316] |
| 154 | $La_2NiO_4$ | $P4_2/ncm$ (138) | *Pc'c'n* | 80 | [89, 317] |
| 155 | $La_2O_3FeMnSe_2$ | *I4/mmm* (139) | *Im'm'm* | 76 | [189, 318] |
| 156 | $Sr_{0.7}Tb_{0.3}CoO_{2.9}$ | *I4/mmm* (139) | *I4'/mmm'* | ~ 300 | [189, 319] |
| 157 | $Sr_{0.7}Er_{0.3}CoO_{2.8}$ | *I4/mmm* (139) | *I4'/mmm'* | ~ 300 | [189, 319] |
| 158 | $Sr_{0.7}Ho_{0.3}CoO_{2.7}$ | *I4/mmm* (139) | *I4'/mmm'* | ~ 300 | [189, 319] |
| 159 | $KMnF_3$ | *I4/mcm* (140) | *I4/mcm* | 86.8 | [24, 274] |
| 160 | β-$Fe_2(PO_4)O$ | $I4_1/amd$ (141) | $I4_1'/am'd$ | 408 | [320-321] |
| 161 | $Co_2(PO_4)O$ | $I4_1/amd$ (141) | $I4_1'/am'd$ | | [322-323] |
| 162 | $Cr_2S_3$ | $R\bar{3}$ (148) | $P\bar{1}$ | | [9] |
| 163 | $MnTiO_3$ | *R3c* (161) | *Cc'* | 28 | [9, 324] |
| 164 | $PbNiO_3$ | *R3c* (161) | *R3c* | 205 | [89, 325-326] |
| 165 | $ZrMnO_3$ | $R3c$ (161) | | | [190] |
| 166 | $GaFeO_3$ | $R3c$ (161) | $Cc'$ | 408 | [189, 327] |
| 167 | $ScFeO_3$ | $R3c$ (161) | $Cc'$ | 360(5) | [189, 328] |
| 168 | $Ho_{0.1}Bi_{0.9}FeO_3$ | $R3c$ (161) | $R3c$ | | [189, 276] |
| 169 | $Ho_{0.05}Bi_{0.95}FeO_3$ | $R3c$ (161) | $R3c$ | | [189, 276] |
| 170 | $K_{1.62}Fe4O_{6.62}(OH)_{0.38}$ | $P\bar{3}1c$ (163) | $P\bar{3}1c$ | | [189, 329] |
| 171 | $MnCO_3$ | $R\bar{3}c$ (167) | $C2/c$ | 31.5 | [89, 330] |
| 172 | $FeBO_3$ | $R\bar{3}c$ (167) | $C2'/c'$ | 348 | [89, 331] |
| 173 | $FeCO_3$ | $R\bar{3}c$ (167) | $R\bar{3}c$ | 38 | [89, 332] |
| 174 | $NiCO_3$ | $R\bar{3}c$ (167) | $C2/c$ | 24.25 | [89, 333] |
| 175 | $CoCO_3$ | $R\bar{3}c$ (167) | $C2/c$ | 18.1 | [189, 334] |
| 176 | $CoF_3$ | $R\bar{3}c$ (167) | $R\bar{3}c$ | 460 | [24, 335-336] |
| 177 | $FeF_3$ | $R\bar{3}c$ (167) | $C2'/c'$ | 394 | [24, 335-336] |
| 178 | $CrF_3$ | $R\bar{3}c$ (167) | | | [190] |
| 179 | $NiF_3$ | $R\bar{3}c$ (167) | | | [190] |
| 180 | $VF_3$ | $R\bar{3}c$ (167) | | | [190] |
| 181 | α-$Fe_2O_3$ | $R\bar{3}c$ (167) | $P\bar{1}$ | 259.1(2) | [9, 152, 337] |
| 182 | α-$Fe_2O_3$ | $R\bar{3}c$ (167) | $C2'/c'$ | 955 | [9, 152, 337] |
| 183 | $Ca_3LiRuO_6$ | $R\bar{3}c$ (167) | $C2'/c'$ | 117.0(8) | [89, 338] |
| 184 | $Sr_3LiRuO_6$ | $R\bar{3}c$ (167) | $C2'/c'$ | 90 | [89, 339] |
| 185 | $Sr_3NaRuO_6$ | $R\bar{3}c$ (167) | $C2'/c'$ | 70 | [339] |
| 186 | $Ca_3LiOsO_6$ | $R\bar{3}c$ (167) | $C2'/c'$ | 117.1(9) | [89, 340] |
| 187 | $LaCrO_3$ | $R\bar{3}c$ (167) | $R\bar{3}c$ | | [189, 260] |
| 188 | $Ca_3Co_{2-x}Mn_xO_6$ (x≈0.96) | $R\bar{3}c$ (167) | $R3c$ | 16.5 | [189, 341] |
| 189 | $La_{0.33}Sr_{0.67}FeO_3$ | $R\bar{3}c$ (167) | $C2/c$ | 200 | [9, 342] |
| 190 | $CoNb_3S_6$ | $P6_322$ (182) | $C2'2'2_1$ | 25 | [9, 93, 132, 343] |
| 191 | $VNb_3Sb_6$ | $P6_322$ (182) | $C2'2'2_1$ | 50 | [24, 344] |
| 192 | $Fe_2Mo_3O_8$ | $P6_3mc$ (186) | $P6_3'm'c$ | 59(3) | [189, 345] |
| 193 | $Co_2Mo_3O_8$ | $P6_3mc$ (186) | $P6_3'm'c$ | 42(3) | [189, 345] |
| 194 | $MnO$ | $P6_3mc$ (186) | | | [190, 346] |
| 195 | $MnSe$ | $P6_3mc$ (186) | | | [347] |
| 196 | $UNiGa$ | $P\bar{6}2m$ (189) | $P\bar{6}'2m'$ | 38 | [189, 348] |
| 197 | $FeS$ | $P\bar{6}2c$ (190) | $P\bar{6}'2c'$ | | [189, 349] |
| 198 | $CsCr_3Sb_5$ | $P6/mmm$ (191) | | | [350] |
| 199 | $BaMnO_3$ | $P6_3/mmc$ (194) | $P6_3'/m'cm'$ | 2.3 | [189, 351] |
| 200 | $CrNb_4S_8$ | $P6_3/mmc$ (194) | $P6_3'/m'm'c$ | | [89, 352] |
| 201 | $Ba_3NiRu_2O_9$ | $P6_3/mmc$ (194) | $P6_3'/m'm'c$ | | [89, 353] |
| 202 | $Ba_5Co_5ClO_{13}$ | $P6_3/mmc$ (194) | $P6_3'/m'm'c$ | 110 | [189, 354] |
| 203 | $MnSe$ | $P6_3/mmc$ (194) | | 120(10) | [347] |
| 204 | $CaMnN_2$ | $P6_3/mmc$ (194) | | | [190] |
| 205 | $EuIn_2As_2$ | *P6₃/mmc* (194) | | | [355] |
| 206 | $CsNiCl_3$ | $P6_3/mmc$ (194) | $C22'2_1'$ | 4.85 | [189, 356] |
| 207 | $CsCoCl_3$ | $P6_3/mmc$ (194) | $P6_3'/m'cm'$ | 20.82 | [189, 357] |
| 208 | $RbCoBr_3$ | $P6_3/mmc$ (194) | $P6_3'/m'cm'$ | 36 | [189, 358] |
| 209 | $CoNb_4Se_8$ | $P6_3/mmc$ (194) | | 168 | [359] |
| 210 | $MnTeLi_{0.003}$ | $P6_3/mmc$ (194) | $C2'/m'$ | 307 | [189, 360] |
| 211 | $MnSe_2$ | $Pa\bar{3}$ (205) | $Pca'2_1'$ | 49 | [189, 361] |
| 212 | $MnSe_2$ | $Pa\bar{3}$ (205) | *Pbca* | | [189, 362] |
| 213 | $Co_2Mo_3N$ | $P4_132$ (213) | $C22'2_1'$ | | [189] |
| 214 | $Ce_4Sb_3$ | $I\bar{4}3d$ (220) | $I\bar{4}'2d'$ | | [189, 363-364] |
| 215 | $Ce_4Ge_3$ | $I\bar{4}3d$ (220) | $I\bar{4}2d$ | ~5 | [189, 365] |
| 216 | $Tb_2C_3$ | $I\bar{4}3d$ (220) | $Fd'd2'$ | 33(4) | [189, 366] |
| 217 | $LiTi_2O_4$ | $Fd\bar{3}m$ (227) | $I4_1'/am'd$ | | [367] |
| 218 | $Bi_2RuMnO_7$ | $Fd\bar{3}m$ (227) | $Fd'd'd$ | 20 | [189, 368] |
| 219 | $Er_2Ru_2O_7$ | $Fd\bar{3}m$ (227) | $I4_1'/am'd$ | 90 | [189, 369] |
| 220 | $La_2NiO_4$ | *Bmab* | | 770 | [317, 370] |
| 221 | $La_2CuO_4$ | *Bmab* | $P_Accn$ | 316(2) | [24, 371] |

## 5. Conclusion

In this review, we explore the intriguing world of AMs, focusing on their physical properties, the various methodologies used to induce altermagnetism, and the promising candidates for altermagnetic materials. As a magnetic phase that has only recently been identified, the study of AMs remains in its foundational stages, and the existence of a real material that exhibits all the predicted features of altermagnetism has yet to be conclusively demonstrated. This nascent field promises a wealth of uncharted physical properties, which may pave the way for groundbreaking applications in technology. Given their unique magnetic behaviors, AMs could revolutionize fields such as data storage, spintronics, and magnetic sensing, offering more efficient and durable alternatives to traditional ferromagnetic and antiferromagnetic materials. The coming decades may witness significant advancements as researchers continue to uncover the full potential of these materials, potentially leading to the development of faster, smaller, and more energy-efficient electronic devices that leverage the unique properties of AMs.

## Acknowledgements

This work is supported by the National Key R&D Program of China (Grant No. 2022YFA1403800, No. 2022YFA1402600, and No. 2020YFA0308800), the National Natural Science Foundation of China (Grant No. 12274027, No. 12234003, No. 12321004 and No. 12347214), and the Fundamental Research Funds for the Central Universities (Grant No.

2024CX06104). Y. M. and L. S. acknowledge support by the DFG-TRR 288-422213477/2 and DFG-TRR 173-268565370/3. We also acknowledge the Joint Sino-German Research Projects (Chinese Grant No. 12061131002 and German Research Foundation, DFG, Grant No. 44880005) for funding.

This review provides an overview of the recent progress in the study of emergent magnetic phase known as altermagnetism. The unconventional properties of altermagnets, including lifted Kramer degeneracy, anomalous and spin transport properties, magneto-optical effects and chiral magnons, along with several avenues to inducing altermagnetism are introduced. Additionally, the two-dimensional and three-dimensional altermagnets discovered so far are summarized.

Ling Bai, Wanxiang Feng*, Siyuan Liu, Libor Šmejkal, Yuriy Mokrousov, Yugui Yao*

Altermagnetism: Exploring New Frontiers in Magnetism and Spintronics

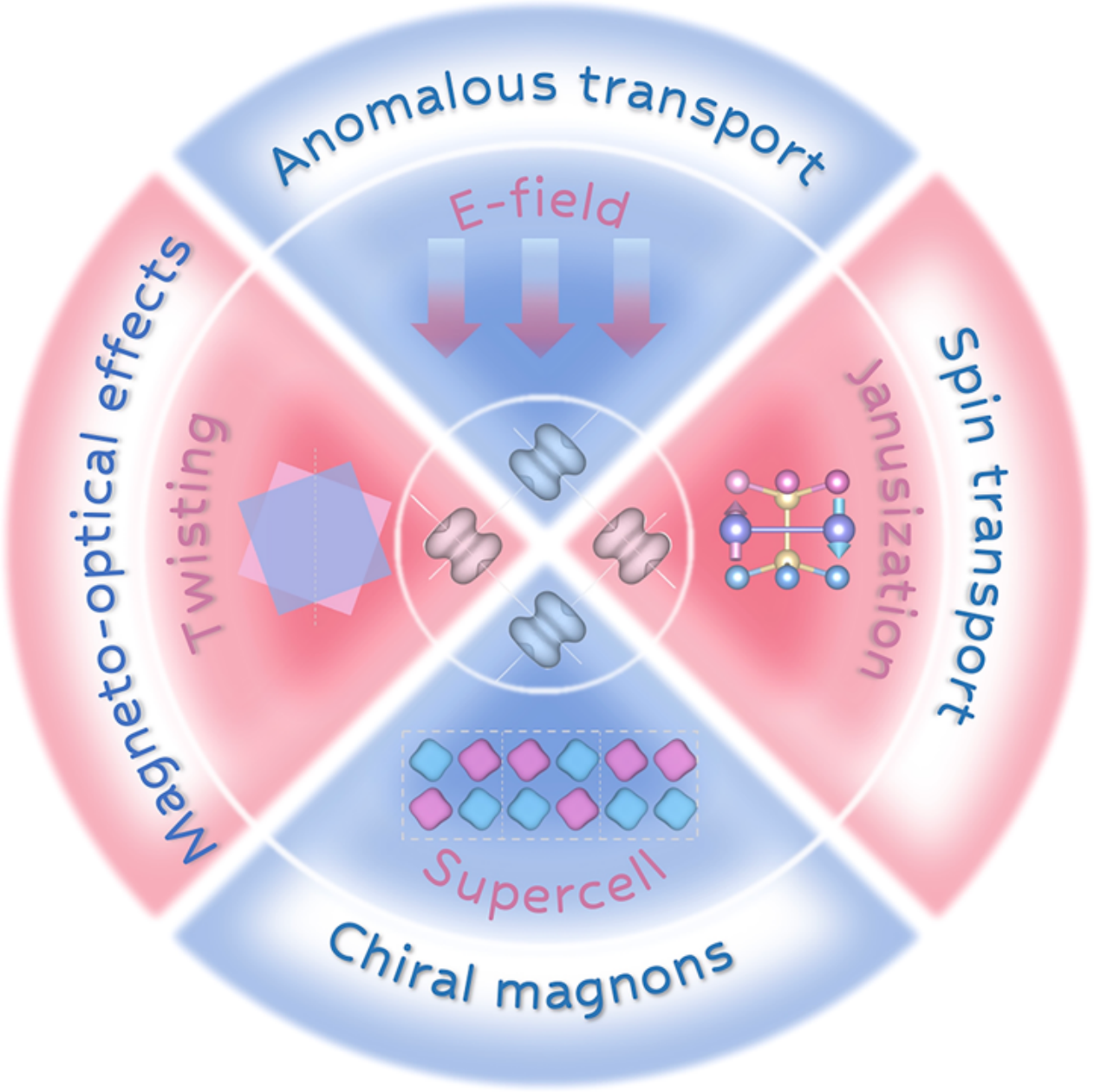